\documentclass[useAMS,usenatbib,twocolumn,floatfix]{mn2e}
\usepackage{natbib}
\usepackage{graphicx}
\usepackage{bm}
\usepackage{amsmath,amssymb}
\usepackage[dvipdfmx]{color}
\usepackage{color}
\usepackage[dvipdfm,
bookmarkstype=toc=true,
 linktocpage=true,
 bookmarks=true
 ]{hyperref}
\usepackage[varg]{txfonts}

\newcommand{\lm}{\ell m}

\def\uv#1{\hat{\mbox{\boldmath $#1$}}}

\def\rref#1{equation (\ref{#1})}
\def\Rref#1{Equation (\ref{#1})}
\def\Fref#1{Fig. \ref{#1}}
\def\Sref#1{Sec. \ref{#1}}
\newcommand{\bvec}[1]{\mbox{\boldmath $#1$}}

\newcommand{\dd}{{\rm d}}

\begin{document}

\title[Reconstruction of Missing Data using IHE]%
{Reconstruction of Missing Data using Iterative Harmonic Expansion}

\author[Nishizawa and Inoue]
{Atsushi  J. Nishizawa$^{1,2}$\thanks{Email: atsushi.nishiza@iar.nagoya-u.ac.jp} 
and Kaiki Taro Inoue $^{3}$\\
$^{1}$ Institute for Advanced Research, Nagoya University, Aichi 464-8602, Japan,\\
$^{2}$ Kavli Institute for the Physics and Mathematics of the Universe (WPI),
  The University of Tokyo, Chiba 277-8583, Japan \\
$^{3}$ Faculty of Science and Engineering, Kindai University, Higashi-Osaka, 577-8502, Japan}
\maketitle

\begin{abstract}
  In the cosmic microwave background or galaxy density maps,
  missing fluctuations in masked regions can be reconstructed from
  fluctuations in the surrounding unmasked regions if the original
  fluctuations are sufficiently smooth.  One reconstruction method
  involves applying a harmonic expansion iteratively to fluctuations
  in the unmasked region.  In this paper, we discuss how well this
  reconstruction method can recover the original fluctuations
  depending on the prior of fluctuations and property of the masked
  region.  The reconstruction method is formulated with an asymptotic
  expansion in terms of the size of mask for a fixed iteration number.
  The reconstruction accuracy depends on the mask size, the spectrum
  of the underlying density fluctuations, the scales of the
  fluctuations to be reconstructed and the number of iterations.  For
  Gaussian fluctuations with the Harrison--Zel'dovich spectrum, the
  reconstruction method provides more accurate restoration than naive
  methods based on brute--forth matrix inversion or the singular value
  decomposition.  We also demonstrate that an isotropic non-Gaussian
  prior does not change the results but an anisotropic non-Gaussian
  prior can yield a higher reconstruction accuracy compared to the
  Gaussian prior case.
\end{abstract}

\section{introduction}
\label{sec:introduction}
After the first data release of the cosmic microwave background (CMB)
temperature fluctuations observed by the Wilkinson Microwave
Anisotropy Probe (WMAP), 
multiple authors reported anomalous signatures, so-called
  ``large--angle anomalies'', in the CMB on large angular scales
\citep{RalstonJain:04,deOliveiraetal:04,Hansenetal:04,Hajianetal:05,
  Moffat:05,LandMagueijo:06,Bernuietal:06,Copietal:07,Eriksenetal:07,
  Monteserinetal:08,Samaletal:09}
which were confirmed recently by the Planck Collaboration
\citep{PlanckXXIII:13, Planck2015:isotropy}.
To date, the origin of these anomalies has not been addressed.
 They may be due to
(a) 
a difference between a priori and posteriori significance
\citep{AurichLustig:10,PontzenPeiris:10,Efstathiouetal:10,Bennettetal:11}
(b) incomplete subtraction of
foreground emissions 
\citep{Abramoetal:06,Cruzetal:11,Hansenetal:12}
(c) 
a
contribution from large--scale structures via the integrated Sachs--Wolfe effect 
\citep{InoueSilk:06,InoueSilk:07,Rassatetal:07,Afsholdietal:09,FP:10,Rassatetal:13,RassatStarck:13,
TomitaInoue:08, SakaiInoue:08,Inoue:12,PlanckXXIII:13}, or kinetic Sunyaev--Zel'dovich 
effect \citep{Peiris:10},
(d) possible systematics from instruments \citep{Hansonetal:10}
(e) incomplete treatment of masking \citep{Kimetal:12,Rassatetal:14}, or
(f) extensions of inflationary models
\citep{Aurichetal:07,Gumrukcuogluetal:07,Rodrigues:08,Bernuietal:08,
Cruzetal:08,Fialkov:10,ZhengBunn:10,Liuetal:13}.

The entire sky cannot be directly observed with a sufficiently high
signal--to--noise (S/N) ratio. A conservative approach involves
masking out the regions where the S/N ratio is low and the signal is
highly contaminated by foreground emissions (e.g., the Zone of
Avoidance) and simply ignoring the data in such regions.  Another
approach involves reconstructing the missing fluctuations in the
masked region based on those outside the masked region and using the
reconstructed data as well.  However, to do so, it is necessary to
make certain assumptions regarding the prior on the property of the
missing fluctuations.  To estimate the power spectrum of the
fluctuations from an incomplete sky, we can use deconvolution
techniques \citep[e.g.][]{Hivon:2002} if we adopt a prior that the
fluctuations are statistically isotropic.  However, to estimate the
density field itself, it is necessary to develop methods that can
reconstruct the phases (if expressed in complex numbers) as well as
the amplitudes of the missing fluctuations.

To reconstruct missing fluctuations on the masked region, it is
necessary to find the inverse of the masking operator.  However, in
general, the mask matrix is singular and therefore is not invertible.
It is necessary to make certain assumptions about the underlying data
such as isotropy and smoothness
\citep{Abrialetal:08,Kimetal:12,BucherLouis:12,Starcketal:13} or their
derivatives \citep{Inoueetal:08} to regularize the inverse operator.
Because the result depends on the choice of the prior, the mutual
robustness of each reconstruction method should be checked.

In this paper, we revisit the iterative harmonic expansion (IHE) for
the regularization of the inverse of a masking operator.  This method
is well known as the Jacobi iterative process and has been applied to
create CMB maps, as reported in the literature \citep{Prunet+00,
  Hamilton03} which is fast and easy to use.  It has already been
implemented in the {\small HEALPIX} package as {\it
  map2alm\_iterative}\footnote{http://healpix.jpl.nasa.gov/}
\citep{Gorski:2005}.  The IHE
method is quite robust against the statistical properties of the
fluctuations.  In this paper, we show that the IHE method does not
require statistical isotropy or Gaussianity for the fluctuation to be
reconstructed. We also demonstrate that the underlying power spectrum
of the fluctuations greatly influences the reconstruction accuracy.
For simplicity, we ignore the noise components in our discussion.
Because our main purpose in the study described herein is to apply the
IHE method to reconstruct the large--angle CMB fluctuations
contaminated by the foreground, this assumption is reasonable.

This paper is organized as follows.  In \Sref{sec:iteration}, we
describe the formulation of the IHE method using the IHE on an
$N$--dimensional unit sphere and provide a verification of the IHE
method based on asymptotic expansion.  In \Sref{sec:result}, we
describe the simulation set we used and show some numerical results to
compare the reconstruction accuracy of the IHE method to that of the
brute--force inversion or the singular value decomposition (SVD)
method. We also discuss the masking effect and the reconstruction
accuracy for different $\ell$ and $m$ modes.  In \Sref{sec:realistic},
we present the application of the IHE method to the CMB sky and
non-Gaussian fluctuations. In \Sref{sec:summary}, we give our
conclusions.

\section{Iterative harmonic expansion}
\label{sec:iteration}

In this section, we describe the IHE method of reconstructing missing
fluctuations on a masked region. In \Sref{ssec:theory}, we formulate
the IHE method for an $N$--dimensional unit sphere.  In
Secs. \ref{ssec:1d} and \ref{ssec:2d}, we discuss the asymptotic
expansion of the mask matrix in terms of the size of a masked region
on a circle and a two dimensional sphere, respectively. In both cases,
we show that the IHE method gives an exact solution in the limit where
the size of the masked region approaches zero when the iteration
number is fixed.
\begin{figure*}
  \includegraphics[width=\linewidth]{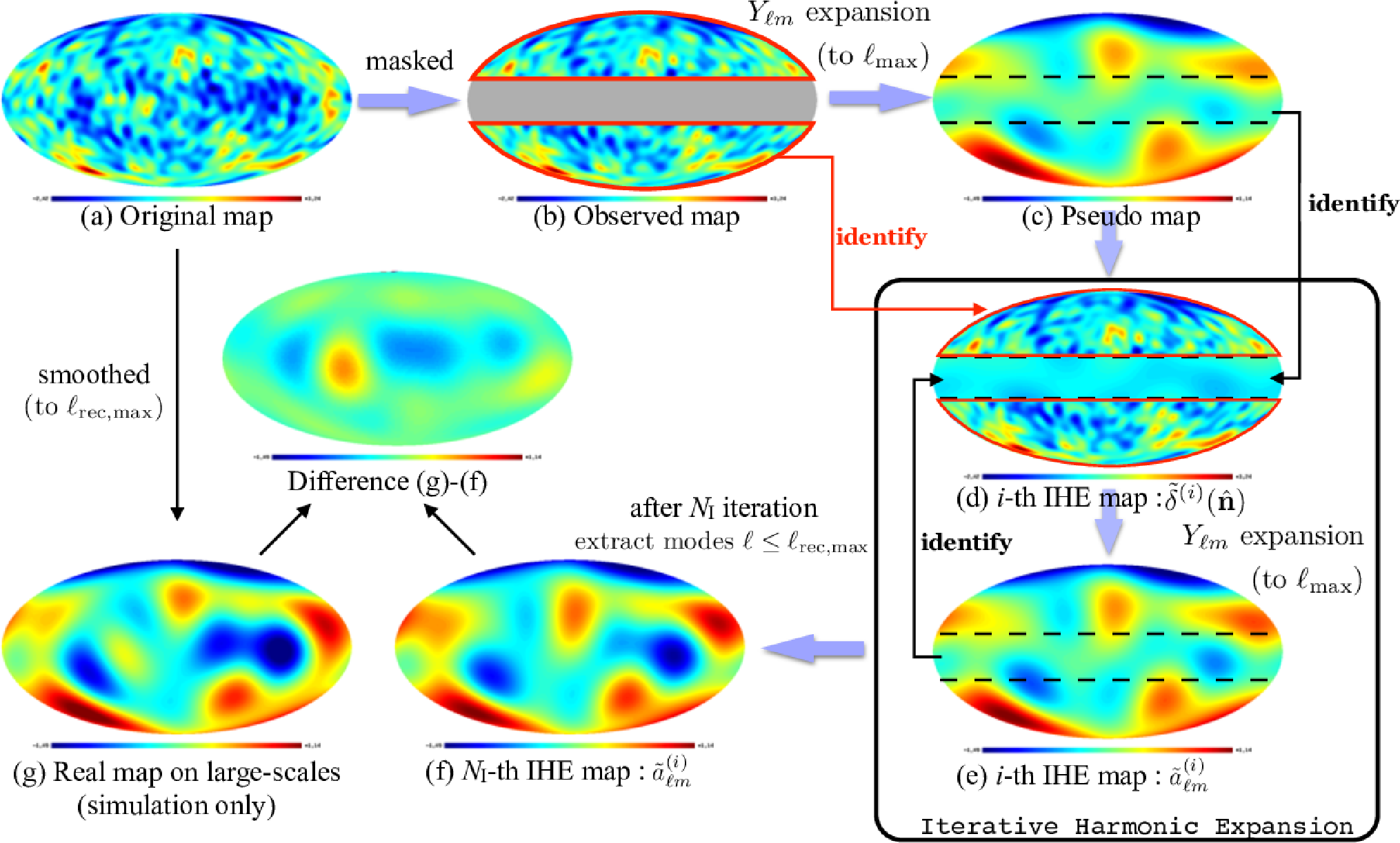}
  \caption{Schematic diagram of the IHE method. (a) 
    Original map including high multipoles. (b)
    Original map in which an azimuthally symmetric region is masked. 
   (c) (b) is expanded in harmonics up to 
     a given $\ell_{\rm max}$. 
 (d) Combination of 
    the original map (b) outside the mask and the pseudo map 
    (c) inside the mask.
    (e) (d) is expanded up to $\ell_{\rm max}$ and
    inversely transformed to obtain the smoothed map. The process
 (d)--(e) is repeated for $N_{\rm I}-1$ time. 
 (f) We obtain an $N_{\rm I}$ times iterated map and modes up to $\ell_{\rm
   rec,max}$ are extracted.
 (g) The ``true'' fluctuations on large--scales up to $\ell_{\rm rec,max}$}.
  \label{fig:diagram}
\end{figure*}

\subsection{Formulation}
\label{ssec:theory}
Suppose a density fluctuation $\delta(\uv{\gamma})$ on a unit
$N$--sphere $\textrm{S}^N$, where $\uv{\gamma}$ represents a unit
vector pointing to a position in $\textrm{S}^N$. The masking function
is defined by
\begin{align}
  \label{eq:masking}
  \delta_{\rm obs}(\uv{\gamma})=W(\uv{\gamma})\delta(\uv{\gamma}),
\end{align}
where $W=0$ inside the masked region and $W=1$ elsewhere.
The fluctuation can be expanded in terms of spherical harmonics $X_i$,
\begin{align}
  \label{eq:harmonics}
  \delta(\uv{\gamma})
  =
  \sum_i a_i X_i(\uv{\gamma}),
\end{align}
where $X_i(\uv{\gamma})$ is a solution of the Helmholtz equation,
$(\Delta_N+k_i^2)X_i=0$ where $\Delta_N$ is the $N$--dimensional
Laplacian on $\textrm{S}^N$ and the $k_i^2$'s are the eigenvalues
defined in ascending order $k_0^2<k_1^2<k_2^2,\cdots$.  We use a
single subscript index $i$ to represent the scale of each mode. For
instance, for a two--dimensional unit sphere $\textrm{S}^2$, the
eigenfunction is $Y_i$, the spherical harmonics and the eigenvalues
are $k_i^2=\ell(\ell+1)$.  The index $i$ of $Y_i(\uv{\gamma})$ is
given by a multipole number $l$ and a magnetic quantum number $m$
$i = \ell^2 + \ell + m + 1$.

The sum in \Rref{eq:harmonics} should be taken over all
$i$--modes. However, in real applications, we can truncate the sum at a
certain scale if the modes on smaller scales are 
not physically
relevant.  The coefficient $a_i$ is called the harmonic coefficient
and can be
obtained from
the inverse transformation of \Rref{eq:harmonics}.
If the fluctuations on the masked region is set to zero, we will
  obtain the
so--called {\it pseudo} harmonic coefficients 
$\tilde{a}_i^{\rm P}$s,
\begin{align}
  \label{eq:pseudo-init}
  \tilde{a}_{i}^{\rm P}
  &=
  \int \dd \Omega~ \delta(\uv{\gamma}) W(\uv{\gamma}) X^*_{i}(\uv{\gamma})
\end{align}
where $\dd\Omega$ is the surface element on $\textrm{S}^N$.
\Rref{eq:pseudo-init} can be written in terms of the true harmonic
coefficients 
$a_{j}^{\rm true}$s,
as
\begin{align}
  \label{eq:pseudo-wlm}
  \tilde{a}_{i} = \sum_j ~ a_{j}^{\rm true} W_{ij},
\end{align}
where $W_{ij}$ is 
the $(i,j)$ component of the mode coupling matrix
due to the masking:
$W_{ij}=\int \dd\Omega~ W X_{j} X_{i}^*$. 

\Rref{eq:pseudo-wlm} implies that 
the true expansion coefficient
$\bvec{a}^{\rm true}$ 
can be obtained
by inverting the mode coupling matrix
$\bvec{W}$.  However, in general, the matrix can be singular and 
non--invertible.
To
regularize a
singular matrix, the IHE scheme can be employed.
The iteration process starts from 
a set of the pseudo harmonic coefficients
\begin{align}
  \label{eq:iteration3}
  \tilde{a}_{i}^{(0)}
  &=
  \tilde{a}_{i}^{\rm P}.
\end{align}
For $n \geq 1$, the $n$--th set of
$\tilde{a}_{i}$s can be
constructed from two maps:
the original map of the unmasked region that was obtained observationally
and the map that was reconstructed
from the inverse transform of the 
$n$--th $\tilde{a}_{i}$s in the
masked region,
\begin{align}
  \label{eq:iteration1}
  \tilde{a}_{i}^{(n)}
  &=
  \int \dd\Omega~ \left[
    \delta(\uv{\gamma}) W(\uv{\gamma}) 
    + \tilde{\delta}^{(n)}(\uv{\gamma}) R(\uv{\gamma})
  \right] X_{i}^{*}(\uv{\gamma}),
\end{align}
where
\begin{align}
  \label{eq:iteration2}
  \tilde{\delta}^{(n)}(\uv{\gamma})
  =
  \sum_i^{i_{\rm max}} ~ \tilde{a}_{i}^{(n-1)}X_{i}(\uv{\gamma}) 
\end{align}
and $R=1-W$ (see also Fig. \ref{fig:diagram}). 
Note that the $n$--th iterated real space map contains
information equivalent to that conveyed by the
$(n-1)$--th harmonic coefficient.
Therefore, we call the $\tilde{\delta}^{(N_{\rm I})}$ as the $N_{\rm
  I}$--th estimator together with $\tilde{a}_{i}^{(N_{\rm I}-1)}$.
In the following,
we assume that fluctuations whose
angular scales are larger than that of the masked region 
are not significantly correlated
with fluctuations smaller than the masked
region.  In that case, the summations in 
Eqs. (\ref{eq:pseudo-wlm}) and (\ref{eq:iteration2})
can be truncated at a certain multipole $i_{\rm max}$, which 
can be
inferred from the size of the mask, as long as we
concern large--angle fluctuations corresponding to multipoles 
$i_{\rm rec,max} \leq i_{\rm max}$. 
In the following,
we sum the
$a_i$'s up to the multipole
$i_{\rm max}$, and we omit the summation symbol when no confusion
arises.

Recursively substituting
\rref{eq:iteration2} into (\ref{eq:iteration1}),
we can
obtain the general formula for
the $N_{\rm I}$--th
iterated harmonic coefficients as a series of $R_{ij}$,
\begin{equation}
  \label{eq:neuman_series}
  \tilde{a}_{i}^{(N_{\rm I}-1)}
  =
  \sum_j^{i_{\rm max}}
  \tilde{a}_{j}^{(0)}
  ( 
  \delta^{K}_{ij}+R_{ij}+R^{2}_{ij}+\cdots+[R^{N_{\rm I}-1}]_{ij}
  ),
\end{equation}
where $\delta^{K}_{ij}$ is the Kronecker delta, 
$R_{ij}=\delta^{K}_{ij}-W_{ij}$, and $R^2_{ij}=\sum_kR_{ik}R_{kj}$ and so on.
This finite series, which is truncated at the
$N_{\rm I}$--th order represents an asymptotic expansion of $W^{-1}$
in terms of the masked region.  As the area of the masked region 
approaches zero, the series converges to the true value $a_i$. As
shown in \Sref{ssec:2d}, the difference between the $N_{\rm I}$--th
estimator and the true $a_i$, 
$\Delta_i^{(N_{\rm I}-1)} = | \sum_j a_j[R^{N_{\rm I}}]_{ij} |$
is of the order of $O(b^{N_{\rm I}})$. 
Therefore,
in the limit of $b \rightarrow 0$, the reconstructed
fluctuations converge to the true fluctuations.
Even if the area of the mask
is finite, the series can converge depending on the following
conditions:
\begin{itemize}
\item{} The mask size: As the volume/area of the mask increases, 
  the rate of convergence slows down
  because the residual matrix $\bvec{R}$
  significantly deviates from zero and the contribution to
  $\Delta_i^{(N_{\rm I})}$ is not negligible.  We shall discuss this
  issue in Secs. \ref{ssec:1d} and \ref{ssec:2d}.
  \item{} The scales we try to reconstruct: The minimum scale
    corresponding to the highest $i$--mode $i_{\rm rec, max}$ to be
    reconstructed should be equal to or less than the scale of
    the $i_{\rm max}$--mode.  The choice of the cut--off scale may change the
    speed of convergence because
    the mode coupling with the high $i$--modes may
    become important. We shall discuss this point in
    \Sref{ssec:lmmode}.
  \item{} The spectrum of underlying density fluctuations: If an
    ensemble averaged density fluctuation has a blue spectrum, the
    effect of mode coupling, especially from high $i$--modes, becomes
    more conspicuous than in the cases with red spectra.
    We shall examine this point in \Sref{ssec:dependence}.
\end{itemize}
Therefore,
\rref{eq:neuman_series} can approximate the underlying true
density fluctuation.  The optimal number of iterations, $N_{\rm I}$,
under the given conditions described above should be evaluated using
Monte--Carlo simulations; this procedure
is discussed later in \Sref{sec:result}.

We can think of
the reconstruction process as a mapping from an
observable to an
estimator,
\begin{equation}
  F_{\rm IHE}: \tilde{a}_i \rightarrow \tilde{a}_i^{(N_{\rm I}-1)}.
  \label{eq:mapping}
\end{equation}
As we briefly mentioned
above and will discuss
in more detail in \Sref{sec:result},
the accuracy of the IHE method depends on the highest mode to be
reconstructed $i_{\rm max}$, the spectrum index $n_s$ of the power
spectrum of $\delta$, the mask size and the number of total iterations
$N_{\rm I}$. Therefore,
we can write $F_{\rm IHE}=F_{\rm IHE}(i_{\rm max}, n_s, \bvec{W}, N_{\rm I})$. 
In \Sref{ssec:real}, we will show that the details of the boundary
shape do not significantly affect the reconstruction accuracy if the
area of the masked sky is similar.

The IHE method can easily be
applied to a sky with an azimuthally
symmetric mask. As shown in \Fref{fig:diagram}, the algorithm is
simple and easy to implement. The IHE method may also work for three
(or higher) dimensional problems provided that the volumes of the
missing regions are sufficiently smaller than the scales of interest.

\subsection{Asymptotic expansion on a circle}
\label{ssec:1d}
In this section and the next section, we demonstrate that the IHE
method gives a finite inversion in the limit that the size of a
masked region approaches zero, even in the case when the size is
finite.  In this section, we consider the reconstruction of the
missing fluctuations in a segment on a circle C with a perimeter
$L=2\pi$ and show that the IHE method is valid in the limit that the
size of the segment converges to zero.

We chose an arbitrary point on C as the origin of the coordinate
$\phi$, which runs
from -$\pi$ to $\pi$.  The mask function is defined as
$W(\phi,b)=0$ for $|\phi|<b/2$ and 1 otherwise, where $b$
characterizes the size of the masked region B: $|\phi|<b/2$.  
Because the eigenfunction $X_m$ corresponding to an eigenvalue $k_m^2=m^2$ is
given by $\exp(-im\phi)$, a density fluctuation defined on C can be
decomposed into discrete Fourier modes as
\begin{align}
  \delta(\phi)
  =
  \sum_{m=0} a_m \exp \left( -i m \phi \right),
\end{align}
where $m$ is an integer with $m=1$ corresponding to the largest Fourier
fluctuation mode on C.  As before, the missing fluctuation in
the masked region B can be
constructed by multiplying the mask function to the
original fluctuation: $\delta_{\rm obs}(\phi, b) = \delta(\phi)
W(\phi, b)$.

The pseudo--estimator for $\delta$, which is equivalent to the 
$N_{\rm I}=1$ IHE is given by
\begin{align}
  \tilde{\delta}^{(1)} (\phi)
  = 
  \sum_{m=0}^{m_{\rm max}} \tilde{a}^{(0)}_m\exp 
  \left( -i m \phi \right),
\end{align}
where the Fourier component $\tilde{a}^{(1)}$ is 
\begin{align}
  \label{eq:a1q}
  \tilde{a}^{(0)}_m
  =
  \int \dd \phi \delta(\phi) W(\phi, b) \exp \left( i m \phi \right).
\end{align}
The mask function can be expressed as
\begin{align}
  W(\phi, b)
  =
  1 - \Theta(\phi+b/2)+\Theta(\phi-b/2),
\end{align}
where $\Theta(\phi)$ is the Heaviside step function.  Then,
\rref{eq:a1q} can be rewritten in terms of a true density $\bvec{a}$
as
\begin{align}
  \tilde{a}^{(0)}_m
  =
  \sum_{m'} ~ a_{m'} W_{mm'},
\end{align}
where $W_{mm'} =\int \dd \phi W(\phi) \exp[i(m-m') \phi]$ corresponds
to the mode coupling matrix in \rref{eq:pseudo-wlm}. Explicitly, it
can be written as
\begin{align}
  W_{mm'}
  =
  \delta^K_{mm'} 
  -
  \frac {2}{m-m'} \sin\left[ \frac{(m-m')b}{2} \right].
\label{wmm'}
\end{align}
In the limit of $b \ll 1$, equation (\ref{wmm'}) can be expanded into
a series as
\begin{align}
  W_{mm'}
  \simeq
  \delta^K_{mm'} - b + \frac{1}{24}(m-m')^2b^3 +\cdots.
\end{align}
Then, the first iterated IHE estimator is
\begin{align}
  \tilde{a}^{(0)}_m
  =
  a_m 
  -
  \sum_{m'} a_{m'} \left[ b - \frac{1}{24}(m-m')^2 b^3 + \cdots
  \right],
\end{align}
which will recover the true density fluctuation when $b \rightarrow 0$.
In a similar manner, the second iterated estimator can be written as
\begin{align} 
  \tilde{a}^{(1)}_m 
  &=
  a_m 
  -
  \sum_{m'm''} a_{m''} \nonumber \\
  &\hspace{2em} \times \left[ -b^2 + \frac{(m-m')^2+(m'-m'')^2}{24}
    b^4 + \cdots \right].
\end{align}
The second iterative estimator also converges to the true
density fluctuation in the limit of $b\rightarrow 0$; however,
it converges faster than the first iterative
estimator because the order of the difference between the estimator
and the original value is $b^2$ rather than $b$. After $N_{\rm I}$
iterations, we obtain
\begin{align}
  \label{eq:1d_asymp}
  \tilde{a}^{(N_{\rm I}-1)}_{m_0}
  &=
  a_{m_0} 
  -
  \sum_{m_1...m_{N_{\rm I}}}
  a_{m_{N_{\rm I}}}
  (-b)^{N_{\rm I}-1} \nonumber \\
  & \hspace{2em}\times
  \left[ b - 
    \left.\frac{1}{24} \right\{ \left. \sum_{n=1}^{N_{\rm I}} (m_{n-1}-m_{n})^2\right\}
    b^{3} + \cdots  \right],
\end{align}
where the difference between the $N_{\rm I}$--th estimator and 
the true fluctuation is of the 
same order as $b^{N_{\rm I}}$.

\subsection{Asymptotic expansion on a two--dimensional sphere}
\label{ssec:2d}
In this section, we consider the IHE reconstruction of missing
fluctuations on a two dimensional sphere.  In what follows, we assume
that the mask is azimuthally symmetric: i.e., $W(\uv{\gamma}, b)=0$
for $|\pi/2-\theta|<b$ and 1 otherwise.  As 
described in \Sref{ssec:1d},
the pseudo-- or 
first IHE estimator can be written as,
\begin{align}
  \label{eq:alm2s-1}
  a^{(0)}_i 
  &=
  \int \!\! \dd\Omega ~ \delta(\uv{\gamma})
  W(\uv{\gamma}, b) Y_{i}(\uv{\gamma}) \\
  \label{eq:alm2s-2}
  &=
  a_i  - \int \!\! \dd\Omega ~ \delta(\uv{\gamma})
  R(\uv{\gamma}, b) Y_{i}(\uv{\gamma}).
\end{align}
Note that 
we have written \rref{eq:alm2s-2}
in terms of
$R=1-W$ rather than $W$ 
to limit the integral range to the
vicinity of $\theta \sim \pi/2$, which 
facilitates analysis of the behavior of the estimator.
Note also that the subscript $i$ denotes a
set of parameters $\ell$ and $m$.  Using equations
(\ref{eq:pseudo-wlm}) and (\ref{eq:iteration3}), we can write
equations (\ref{eq:alm2s-1}) and (\ref{eq:alm2s-2}) as
\begin{align}
  \label{eq:asymp_1}
  a_i^{(0)}
  =
  a_i  -
  \sum_j a_j R_{ij}
\end{align}
where $R_{ij}$ is the residual mask matrix,
\begin{align}
  R_{i_1 i_2} = \sum_{i_3} s_{i_3} T_{i_1 i_2 i_3},
\end{align}
and $s_{i}$ denotes the harmonic expansion of the mask residual,
$s_i = \int \dd\Omega R(\uv{\gamma},b) Y_i^*(\uv{\gamma})$.
The matrix $T$ is explicitly given in \cite{Hivon:2002} as
\begin{align}
  \label{eq:wij-pixel}
  T_{i_1 i_2 i_3}
  &=
  \int \! \dd\Omega~ Y_{i_3} Y_{i_2}^{*} Y_{i_1} \nonumber \\
  &=
  (-1)^{m_2}
  \left[
    \frac{(2\ell_1+1)(2\ell_2+1)(2\ell_3+1)}{4\pi}
    \right]^{1/2} \nonumber \\
  & \hspace{0.5cm}\times
  \left(
  \begin{array}{ccc}
    \ell_1&\ell_2&\ell_3 \\
    0     &0     &0
  \end{array}
  \right)
  \left(
  \begin{array}{ccc}
    \ell_1&\ell_2&\ell_3\\
    m_1   &-m_2  &m_3
  \end{array}
  \right),
\end{align}
where $(:::)$ is the Wigner 3$j$--symbol.  Note that the subscript
$i_j$ depends only on $\ell_j$ and $m_j$,
i.e. $i_j=\ell_j^2+\ell_j+m_j+1$.

For an azimuthally symmetric mask, $s_i$ can be analytically integrated and
expanded in terms of $b (\ll 1)$ as
\begin{align}
  s_i
  &=
  \int_0^{2\pi} \!\! d\phi \int_{\pi/2-b}^{\pi/2+b} \!\! d\cos \theta ~
  Y_i(\theta, \phi=0) \nonumber \\
  &=
  2\pi 
  \sqrt{\frac{(2\ell+1)\cos^2(b)}{4\pi}}
  \frac{P_{\ell 1}[\sin(b)]-P_{\ell 1}[\sin(-b)]}{\ell(\ell+1)} \\
  &\simeq
  c_{i}^{(1)} b + c_{i}^{(3)}b^3 +{\mathcal O}[b^4],
\end{align}
where the explicit form of the $c_{i}$'s is given by
\begin{align}
  c_{i}^{(1)}
  &=
  2\sqrt{(2\ell+1)\pi} P_{\ell 0}(0),
  \\
  c_{i}^{(3)}
  &=
  -\frac13 \sqrt{(2\ell+1)\pi} 
  [
  P_{\ell 0}(0) - P_{\ell 2}(0)
  ],
\end{align}
where $P_{\ell m}$ is the associated Legendre function.  Because
$s_i$ is
the first order of $b$ and the matrix $R_{ij}$ is a linear combination of
the $s_i$s, the lowest order of $R_{ij}$ is $b$.  Therefore,
the leading order of the
difference between the first IHE estimator and the true
fluctuation, 
$\Delta^{(0)}_i \equiv |a_i^{(0)} - a_i|$,
is also $b$.  For
the second iterated IHE estimator, we have
\begin{align}
  \label{eq:asymp_2}
  a^{(1)}_i
  =
  a_i - \sum_{j,k} a_k R_{kj}R_{ji},
\end{align}
and, therefore, $\Delta^{(1)}_i = O(b^2)$.  As in the one--dimensional
case, for a given $N_{\rm I}$, we expect that,
$\Delta^{(N_{\rm I}-1)}_i = O(b^{N_{\rm I}})$.  However, note that
$\Delta^{(N_{\rm I})}_i$ scales to $0$ as $b \rightarrow 0$ but does
not approach $0$ as $N_{\rm I} \rightarrow \infty$ due to the mode
coupling between the different multipole modes, as shown in
equation \eqref{eq:1d_asymp}. Instead, it converges to the one for the
direct inversion as illustrated later in \Fref{fig:svd}.

\begin{figure}
  \includegraphics[width=\linewidth]{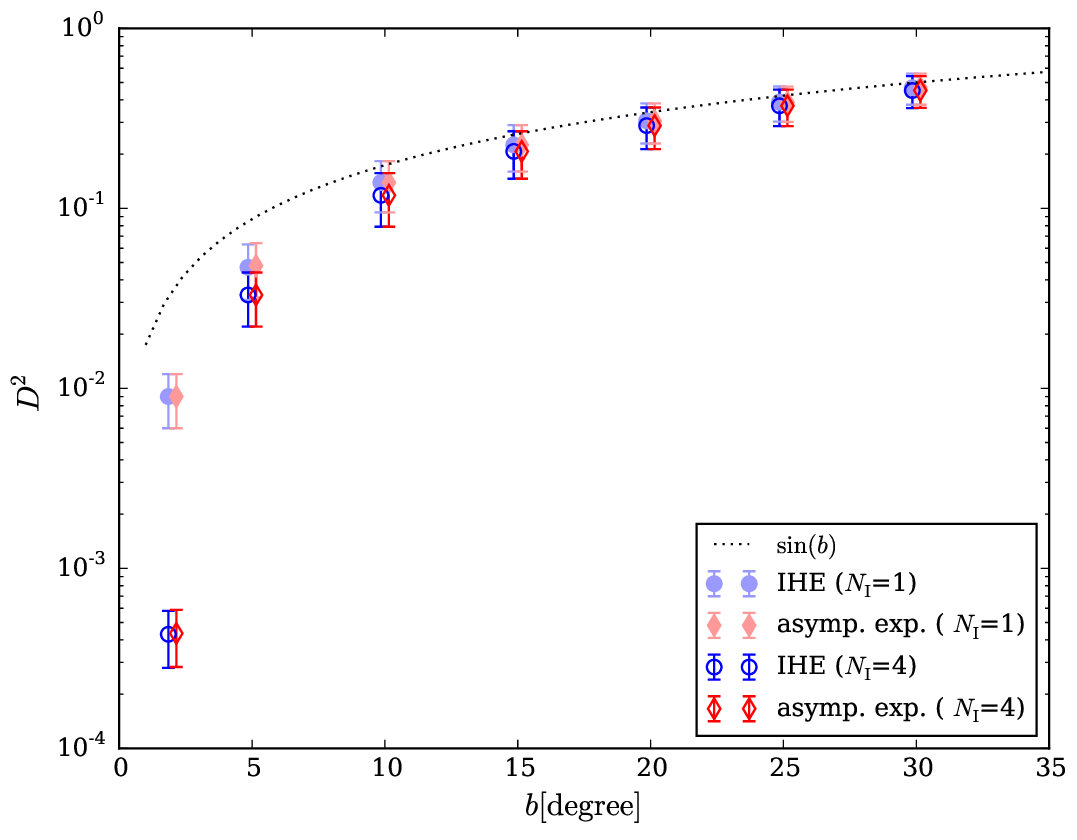}
  \caption{
    Accuracy of fluctuations reconstructed using IHE method and
    truncated asymptotic expansion. Filled and open circles
    represent the $D^2$ accuracies of fluctuations reconstructed
    using IHE method and filled and open diamonds are fluctuations
    obtained using the asymptotic expansion defined in
    \rref{eq:asymp_1}, with filled and open symbols denoting 1 and 4
    iterations, respectively. Also shown with dotted line is the
    $\sin(b)$ curve which corresponds to the expected $D^2$ accuracy
    when masked region is filled with zeros.
    \label{fig:asymptotic}}
\end{figure}

\Fref{fig:asymptotic} presents the $D^2$ accuracy of the IHE map
reconstruction as a function of the angular size of the masked region,
which is compared to the accuracy of the truncated asymptotic
expansion in \rref{eq:asymp_1}.  To compute the $D^2$ accuracies we
used the fiducial set of simulations described in Section
\ref{ssec:data} with various mask sizes.  The circles and diamonds
represent the $D^2$ accuracies defined by \rref{eq:l2norm_l} for the
IHE reconstruction and the truncated asymptotic expansion defined by
\rref{eq:asymp_1}, respectively with filled symbols denoting 1
iteration and open symbols corresponding to 4 iterations.  One can see
that the filled and open circles and diamonds agree very well,
suggesting that the convergence of the IHE method can be numerically
verified.  Furthermore, we expect from \rref{eq:asymp_2} that the
$D^2$ accuracy improves as the number of iterations increases.  This
tendency is more clearly observable at smaller $b$.  In the following,
we discuss the reconstruction of the missing fluctuations on a
two--dimensional sphere;however, in general, the reconstruction method
can be used for an $N$--dimensional sphere.

\section{Results}
\label{sec:result}
In this section, we describe the numerical simulation that we
performed to test the IHE method and present the definition of the
L$^2$ norm that we used to investigate the reconstructed fluctuations
in \Sref{ssec:data}.  Then, we present comparisons of the IHE method
with other inversion methods using the brute--force inversion or the
singular value decomposition (SVD) in \Sref{ssec:svd}. We describe the
fluctuation conditions that can be successfully reconstructed by using
the IHE method in \Sref{ssec:dependence}.  In \Sref{ssec:lmmode},
we discuss the accuracy for each $\ell$ and $m$ mode.

\subsection{Simulations}
\label{ssec:data}
To assess the reconstruction accuracy, we generated $10^3$ random
realisations of an isotropic Gaussian density field in the sky.  We
used the code {\it synfast}, which is publicly available as a package
in {\small HEALPIX} to generate random Gaussian maps.
First, we used the Harrison--Zel'dovich spectrum as the input power
spectrum. It gives an angular power spectrum
$C_{\ell} \propto \ell^{n_s}$, where $n_s=-2$ on large angular scales,
in the Einstein de--Sitter universe, which corresponds to the
Sachs--Wolfe plateau of the CMB power spectrum.  We set the monopole
power to zero because it is a uniform value over the unmasked sky and
can be subtracted out before reconstruction.  As a simple model of the
zone of avoidance, we considered an azimuthally symmetric mask with
$W(\uv{\gamma})=0$ at $|\pi/2-\theta|<b$ and $W=1$ otherwise with
$b=20^{\circ}$.  We used pixels that were sufficiently smaller than
the size of the mask and the reconstructed fluctuation scales to
reduce errors due to the pixelization effect.  We adopt the Healpix
resolution $N_{\textrm{side}}= 1024$ (the total number of pixels in
the entire sky is
$N_{\textrm{pix}} = 12\times N_{\textrm{side}}^2\simeq 1.2\times 10^7$),
where the pixel size corresponds to $3.4$ arcmin. The input power
spectrum should be truncated at scales $\ell_{\rm cut}$ sufficiently
small compared to the reconstructed scales $\ell_{\rm max}$,
i.e. $\ell_{\rm max} \ll \ell_{\rm cut}$.  In our simulated maps, we
set $\ell_{\rm cut}=30$, which is a scale sufficiently smaller than
the mask size.  We also conducted the same analysis with
$\ell_{\rm cut}=100$; however, the result was unchanged.  We use this
set of fiducial simulations unless otherwise stated.  In addition to
this fiducial set, we also generated the same simulation set for the
input spectrum indices $n_s=0$ and $n_s=-4$.

The deviation from the original fluctuations can be measured from the
ratio of the L$^2$ norms denoted by $D^2$,i.e.  the squared difference
between the density fluctuations of the original and reconstructed
maps divided by the squared density fluctuations of the original map:
\begin{align}
  \label{eq:l2norm_l}
  D^2
  \equiv
  \frac{\sum_{i}^{N{\rm pix}} [\delta_{\rm rec}(\uv{\gamma}_i)-
    \delta_{\rm true}(\uv{\gamma}_i)]^2}
  {\sum_{i}^{N{\rm pix}} \delta_{\rm true}^2(\uv{\gamma}_i)}
\end{align}
where $\delta_{\rm rec}$ and $\delta_{\rm true}$ describe the reconstructed and
the original density fluctuations up to the highest multipole:
\begin{align}
  \label{eq:delta_true}
  \delta_{\rm true}(\uv{\gamma})
  &\equiv
  \sum_{i=0}^{i_{\rm rec, max}} a_i^{\rm true} Y_i (\uv{\gamma}), \\
  \label{eq:delta_est}
  \delta_{\rm rec}(\uv{\gamma})
  &\equiv
  \sum_{i=0}^{i_{\rm rec, max}} a_i^{(N_{\rm I})} Y_i (\uv{\gamma}).
\end{align}

\subsection{Comparison with $W^{-1}$ and SVD}
\label{ssec:svd}
\begin{figure}
  \begin{tabular}{l}
    \includegraphics[width=\linewidth]{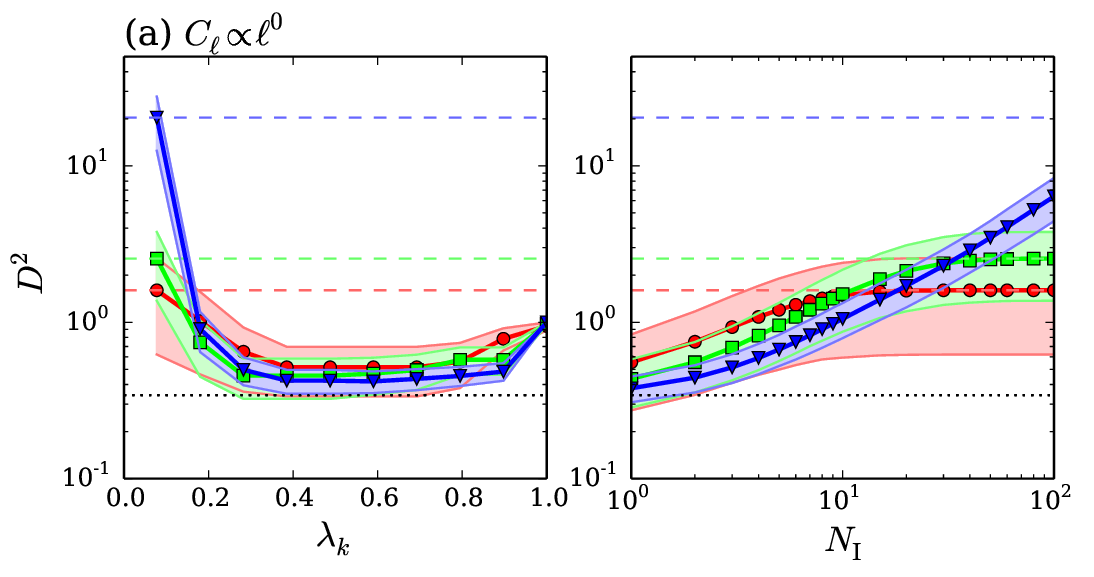}\\
    \includegraphics[width=\linewidth]{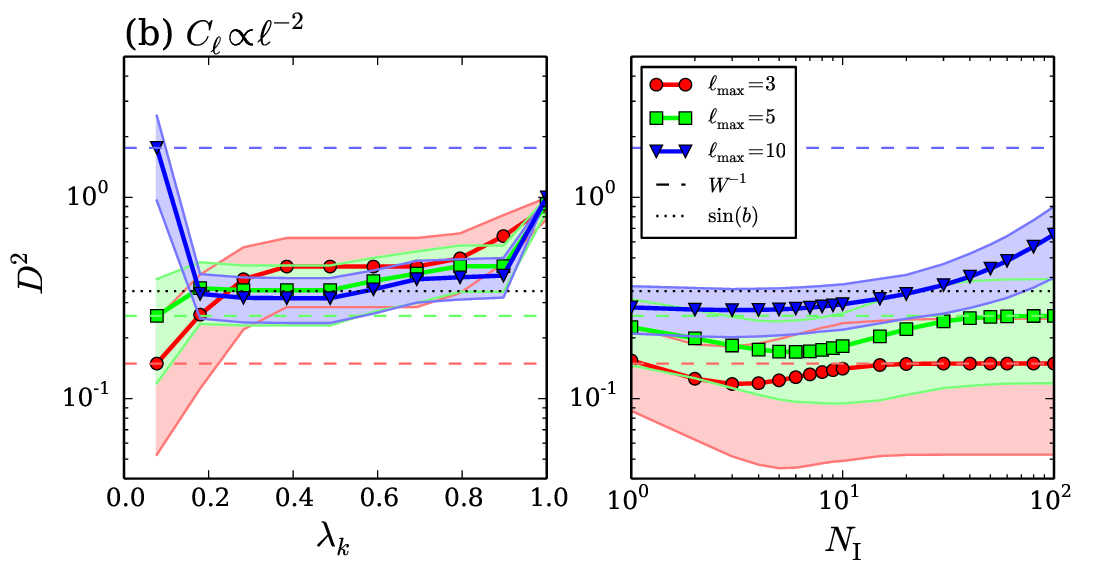}\\
    \includegraphics[width=\linewidth]{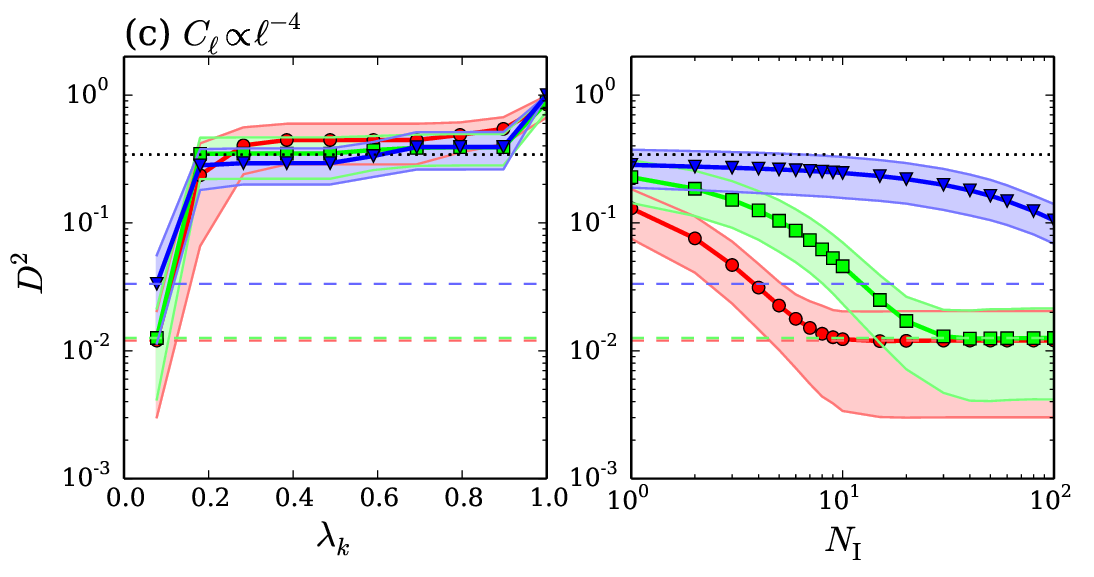}\\
  \end{tabular}
  \caption{
    (Right) $D^2$ accuracy versus iteration number.  
    Panels (a), (b) and (c) show the different 
    underlying density fluctuations that have power--law
    spectra of $C_{\ell} \propto \ell^{0}$, $\ell^{-2}$ and 
    $\ell^{-4}$, respectively. We chose 
    $\ell_{\rm max}=\ell_{\rm rec,max}$.　
    For an $n_s=0$ spectrum, IHE, SVD, and brute--force
    inversion worsen reconstruction results.
    However, for $n_s=-2$, there exists an optimal number of
    iterations near $N_{\rm I}=10$ that minimizes $D^2$
    and depends on $\ell_{\rm max}$. For $n_s=-4$ spectrum, 
    $D^2$ converges to a few percent accuracy after 
    sufficient number of iterations.In all cases, after a 
    sufficient iterations, $D^2$ converges to value 
    obtained by direct inversion. (Left) SVD method results 
    shown for comparison versus eigenvalue threshold 
    $\lambda_k$.
    We also show the $\sin(b)=0.34$ curves where masked 
    region is filled with zeros. For $n_s=0$, ``doing nothing'' 
    is optimal.
    \label{fig:svd}
  }
\end{figure}

Using \rref{eq:pseudo-wlm}, we can compare 
the IHE method to the brute--force inversion of the matrix $\bvec{W}$ 
and the SVD
\citep{Efstathiou:2004}.
If the mode coupling matrix $\bvec{W}$ is invertible, we obtain 
a unique solution for
the underlying density fluctuation within the mask.
However, if the matrix $\bvec{W}$ contains some eigenvalues 
that are close to zero, where the matrix is close to being singular,
the inversion causes large
errors.  In such cases, we can remove the singularities
by replacing the small eigenvalues with zero; this procedure is
  called the SVD method.
Note that the order of the matrix $\bvec{W}$ is closely
  related to the singularity of $\bvec{W}$.  If we take 
the order of
  $\bvec{W}$ to be sufficiently large
to include most of the modes that are
  coupled to $a_i$, the inverse of $\bvec{W}$, if it exists, gives an accurate
  solution of $a_i$. However, the matrix inversion 
  is hampered by the ill-posed nature of the inversion  
  due to singularities.  
  Conversely,
if we truncate the order
  of $\bvec{W}$ at a sufficiently small
  scale $\ell_{\rm max}$, the matrix is
  invertible; however, 
  the inversion matrix gives
a biased solution of $a_i$ because
  the possible couplings from higher modes are
discarded by
  the $\ell_{\rm max}$ truncation. Here we define
  $\bvec{W}$ as an $i_{\rm max} \times i_{\rm max}$ matrix.  
As the first step
of the SVD method, we decompose the matrix $\bvec{W}$ 
into three matrices:
\begin{equation}
  \bvec{W}
  =
  \bvec{U} ~\bvec{\Sigma} ~\bvec{V}^{\dagger},
  \label{eq:svd}
\end{equation}
where $\bvec{V}$ and $\bvec{U}$ are $i_{\rm max}\times i_{\rm max}$
unitary matrices and the
superscript $\dagger$ denotes the Hermitian
conjugate.
$\bvec{\Sigma}$ is the diagonal matrix, which consists of
the eigenvalues of $\bvec{W}$, called the singular values.  The
order of the eigenvalues in $\Sigma$ is arbitrary; however,
they are
arranged in a descending order so that the decomposition is determined
uniquely. Let the $k$--th eigenvalue be $\lambda_k$, and the
eigenvalues smaller than $\lambda_k$ be
zero.  Then, the pseudo--inversion of $W$ can be
written in terms of $\Sigma^{+}$, the rank--$k$
diagonal matrix that consists of the reciprocal of the non--zero
eigenvalues that are equal to or larger than $\lambda_k$:
\begin{equation}
  \bvec{W}^{+}
  =
  \bvec{V} ~\bvec{\Sigma}^{+} ~\bvec{U}^{\dagger},
  \label{eq:svd_inv}
\end{equation}
which gives
\begin{equation}
  a^{\rm est}_i
  =
  \sum_j \tilde{a}_j W^{+}_{ij}.
\end{equation}
The choice of the threshold $\lambda_k$ is not trivial
and should be
carefully determined {\it a priori} because the mapping $F_{\rm SVD}$
depends on various factors including the threshold $\lambda_k$, i.e.
$F_{\rm SVD}=F_{\rm SVD}(\ell_{\rm max}, n_s, {\boldsymbol W},
\lambda_k)$.  In this context,
a {\it singularity} can only be
defined in terms of $\lambda_k$.  More specifically, the mapping
$F_{\rm SVD}$ is contaminated by the eigenvalues that
are close to
being {\it singular} if $\partial D^2 / \partial \lambda_k < 0$. In
such cases, the threshold of the eigenvalues should be increased 
to enable more accurate reconstruction.  The brute--force inversion
corresponds to $F_{\rm inv}=F_{\rm SVD}(\ell_{\rm max}, n_s,
{\boldsymbol W}, \lambda_k=0)$.
The optimal choice of $\ell_{\rm max}$ should depend on 
$n_s$, $\bvec{W}$ and even on $\ell_{\rm rec,max}$. 
Here we simply set $\ell_{\rm max}=\ell_{\rm rec,max}$.

In the left panels of \Fref{fig:svd}, $D^2$ is depicted as a function
of the minimum non--zero eigenvalue $\lambda_k$.  The solid lines
represent the different maximum multipoles to be reconstructed,
$\ell_{\rm max}=3,5$ and $10$.  The three different rows show the
different input power spectra, which will be discussed in detail in
\Sref{ssec:dependence}.  Let us focus on the middle left panel. For
$\ell_{\rm max}=3$ and $5$, $D^2$ increases monotonically as the
threshold increases. Thus, $F_{\rm SVD}$ is not contaminated by
singular eigenvalues because $\partial D^2/\partial \lambda_k > 0$.
On the other hand, Conversely, for $\ell_{\rm max} = 10$, $D^2$ has a
minimum near $\lambda_k\sim 0.4$.  Therefore, eigenvalues smaller than
$\sim 0.4$ may be the contaminants of the mapping $F$, and we can
better estimate the original density fluctuations when we limit the
eigenvalues to $\lambda < 0.4$.  In the right panel of \Fref{fig:svd},
we show $D^2$ for the IHE method as a function of the number of
iterations.  The $D^2$ has a minimum near $N_{\rm I}< 10$ that depends
on the maximum multipole.  After a sufficient number of iterations,
$D^2$ converges to the brute--force inversion results, which are shown
as the horizontal dashed lines in \Fref{fig:svd}, i.e.
$F_{\rm IHE}(N_{\rm I}\rightarrow \infty) = F_{\rm inv}$.  The IHE
method with a certain finite number of iterations is therefore always
more accurate than the SVD method.  However, in practice, we should
know the optimal number of iterations a priori.  This number depends
on the mask size, the maximum multipole to be reconstructed and the
underlying spectrum of the density fluctuation. We can estimate the
optimal iteration number using Monte--Carlo simulations.  In
\Sref{ssec:dependence} we will see how the result changes for
different types of underlying power spectra.

Note that the computation time for estimating
$\tilde{a}^{(N_{\rm I})}$ can be significantly reduced if we use
\rref{eq:iteration1} instead of \rref{eq:neuman_series}.  The reason
is as follows.  Because the rank of the matrix $R_{ij}$ is on the
order of $\ell_{\rm max}^2$, it costs $N_{\rm pix}\ell_{\rm max}^4$
according to \rref{eq:wij-pixel}, and the matrix algebra of
\rref{eq:neuman_series} costs $\sim N_{\rm I} \ell_{\rm max}^6$
computations.  If one uses (\rref{eq:iteration1}, it would be on the
order of $N_{\rm I} N_{\rm pix}\ell_{\rm max}^2$, where $N_{\rm I}$ is
the number of iterations, $N_{\rm pix}$ is the number of pixels that
tile the sky and $\ell_{\rm max}$ is the maximum multipole to be
reconstructed.  For example, for given $\ell_{\rm max}=10$,
$N_{\rm pix}=12\times 1024^2$ pixels and five iterations, the
computation time will be reduced by a factor of $\sim 25$.

\subsection{Statistical properties}
\label{ssec:isotropy}
The statistical isotropy of the CMB map
\citep[e.g.][]{Planck2015:isotropy} and its Gaussianity
\citep[e.g.][]{Planck2015NG} has previously been examined.  When the
map reconstruction method is applied to the CMB map, the statistical
properties of the CMB temperature fluctuations should remain
unchanged.  Using our fiducial simulation set
($n_s=-2, \ell_{\rm rec,max}=5$ and $b=20^{\circ}$), which is
described in \Sref{ssec:data}, we discuss how the IHE method affects
the underlying statistical properties via the reconstruction.

\subsubsection{Statistical isotropy}
\label{sssec:isotropy}
\begin{figure*}
  \includegraphics[width=\linewidth]{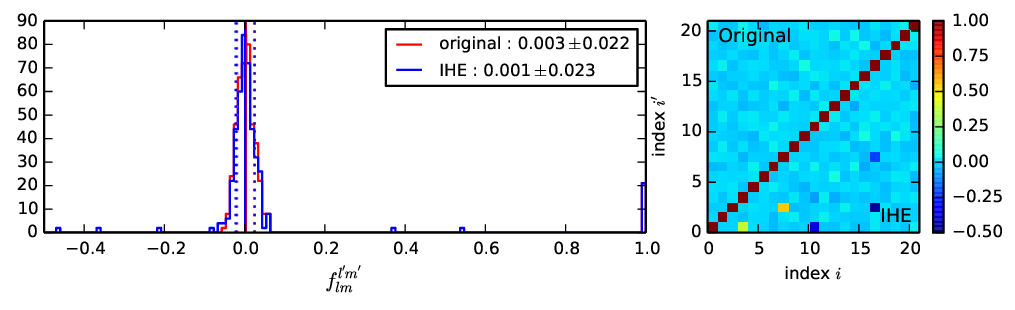}
  \caption{
    ({\it Left}) Distribution of $f$ over different
    modes. Dashed vertical lines show the $1\sigma$ regions.
    Inset values are the mean and the standard deviation.
    ({\it Lower Right}) 
    Components
    of $f$ in
    original
    map (upper left triangle) and the IHE reconstruction with
    $N_{\rm ite}=6$ (lower right triangle).
  \label{fig:isotropy}}
\end{figure*}

First, the statistical isotropy of a fluctuation can be measured
from the ratio between the off-diagonal and corresponding 
diagonal terms in the correlation matrix of the expansion coefficients 
\citep{Inoue2000}:
\begin{align}
  f_{\lm}^{\ell' m'}
  \equiv
  \frac{ \langle a_{\lm} a_{\ell'm'}^* \rangle }%
  { \sqrt{\langle |a_{\lm}|^2 \rangle \langle |a_{\ell'm'}|^2 \rangle} },
\end{align}
where the ensemble average is taken over 1000 realizations. For
statistically isotropic fluctuations, we expect that the off-diagonal
terms satisfy $f \ll 1$. To visualize the matrix, we contract the
four--dimensional subscript into two--dimensional ones as before.
Figure \ref{fig:isotropy} illustrates the distribution of
$f_{\lm}^{\ell'm'}$.  The right panel shows the values for each
element of $f_{\lm}^{\ell'm'}$ calculated from the input map, which
contains only the largest modes, $\ell < 6$.  Because the matrix
component is symmetric, we present the $f_{\lm}^{\ell'm'}$ elements
for the original fluctuation in the upper left triangle matrix and
those for the IHE reconstructed method in the lower right triangle
matrix.  The original map clearly satisfies statistical isotropy.  The
left panel shows a histogram of $f$ with the inset values being
the mean and the standard deviation.  There are a few modes with large
reconstructed fluctuations; however, $f$ is statistically consistent
with zero within $1\sigma$.

\subsubsection{Probability distribution}
\label{sssec:pdf}
Next, we describe
how the probability distribution function 
(PDF)
of the
  fluctuations is modified by a reconstruction using the
Kolmogorov-Smirnov 
(KS) test.
The KS test enables us to
discriminate 
the difference 
between the two underlying 
PDFs
in a non-parametric manner, or
the difference 
between the PDFs taken from a single sample of an
  assumed function.

First, we 
conducted
two sample KS tests using a fiducial set of 1000 random Gaussian
simulations. Let $x_i$ be the pixel value of a certain realization of
an original simulated map at the position of the $i$-th pixel
and
$y_i$ be the corresponding reconstructed pixel value. Then, we define
the cumulative distribution functions (CDFs) for these two data as
\begin{align}
  & F_n(x) = \frac{1}{n} \sum_i^n U(x_i-x), \nonumber \\
  & F_m(x) = \frac{1}{m} \sum_i^m U(y_i-x),
\end{align}
where $U(x)$ is the unit step function and $n=m=N_{\rm pix}$.
In practice,
we divided the sky into
$N_{\rm pix}$ pixels of the
Healpix with a resolution of $N_{\rm side}=32$. The KS statistic is the supremum of the difference between
the two measured CDFs:
\begin{align}
  D_{nm}
  =
  \sup_x \left| F_n(x) - F_m(x) \right|.
\end{align}
Based on
the hypothesis that the reconstructed PDF is consistent with the
original one, the probability that the statistic has a value larger than
$\displaystyle \chi_{nm}=\sqrt{\frac{nm}{n+m}} D_{nm}$ is
\begin{align}
  \Psi(\chi_{nm} < \chi )
  =
  1-\sum_{j=-\infty}^{\infty} (-1)^j \exp[ -2j^2\chi^2 ].
\end{align}
To detect the difference in the shape of the PDF, we 
consider the degeneracy due to different statistical measures. 
To do so, we rescaled
the data as $y'=(y-m)/\sigma$, where $m$ is the mean
and $\sigma$ is the standard deviation. 
We found
that $\langle \Psi \rangle=0.104$ for the IHE reconstructions
with six
iterations, where an ensemble average
was taken over 1000 random realizations. 
Thus,
the PDFs of the reconstructed fluctuations and
the original fluctuations 
were consistent within the $2\sigma$ level.

However, the significance
decreased once we took into account the spatial correlation of the 
fluctuations on the pixels. Such cases have been
observed in cosmological and astronomical signals \citep{Olea2008}.
The covariance of the fluctuations in the
pixels in the sky is given by
\begin{align}
  \langle x(\uv{n}_i) x(\uv{n}_j) \rangle
  =
  C(\theta_{ij}),
\end{align}
where $\uv{n}_i$ points to the sky position of the $i$-th pixel, and
$\cos \theta_{ij} = \uv{n}_i \cdot \uv{n}_j$.
We constructed the
covariance matrix 
directly 
from our 1000 realizations.
If the PDF of the fluctuations is 
Gaussian
or,  more generally, is
fully described in the quadratic form of the data, we can diagonalize the covariance
matrix as
\begin{align}
  M
  = O^{T} C O,
\end{align}
where $M={\rm diag}[\lambda_1, \lambda_2, \cdots, \lambda_n]$,
$\lambda$'s are eigenvalues rearranged in
descending order and
the matrix $O$ consists of $n$ eigenvectors.
If the 
fluctuations are highly spatially
correlated, the rank of the matrix $M$
is always less than $n$; therefore, 
for $i> r_M$, $\lambda_i=0$, where 
$r_M$ is the rank of $M$.
Then the de-correlated data is shrunk
such that
\begin{align}
  x_i \rightarrow x_i'
  =
  \frac{1}{\lambda_i}\left( O^T x \right)_i, ~~ {\rm for} ~ i\le r_M.
\end{align}
With this de-correlated dataset, we found
that a KS test yielded
$\langle \Psi \rangle=0.64$ for the IHE reconstructions, 
still fully consistent with the original PDF.

Finally, we 
applied a single-sample
KS test to 
determine whether the
PDF of the data 
was Gaussian.
Using the rescaled and de-correlated data for this test, we 
found that $\langle \Psi \rangle=0.57$ and $0.58$ for the original and
IHE reconstructed fluctuations, respectively. 
Therefore,
we concluded that the original map was
fairly consistent 
with a Gaussian distribution
and that the IHE reconstruction
did not change the underlying statistical  properties significantly.

\subsection{Power spectrum reconstruction accuracy}
\label{ssec:powerspectrum}
In 
\Sref{sssec:pdf}, 
we demonstrated
that the IHE reconstruction does
not  significantly
change the statistical properties
of the map.
In this section, we will show
how a two-point statistic, the power
spectrum, is affected or recovered
using the IHE method.
Again, we use our fiducial simulation set to access the reconstruction.
The power spectrum of the $\alpha$-th realization map
can be estimated to be
\begin{align}
  \hat{C}_{\ell}^{\alpha}
  =
  \frac{1}{2\ell+1} \sum_m \left| a_{\lm}^{\alpha} \right|^2,
\end{align}
where the true power spectrum
can be
estimated from the arithmetic mean
over 1000 realizations, $C_\ell = \langle \hat{C}_{\ell} \rangle =
\sum_{\alpha} \hat{C_{\ell}^{\alpha}} / 1000$ with 
a variance of
$\sigma^2_{\ell} = \sum_\alpha (\hat{C}_{\ell}^{\alpha} - C_{\ell})^2 / 999$.
To quantify the discrepancy between the power
spectra
of the reconstructed and original fluctuations,
we use the relative difference summed over
the multipoles,
\begin{align}
  \delta C
  = 
  \sum_{\ell}
  \left(
  \frac{C^{\rm rec}_{\ell}}{C^{\rm org}_{\ell}}-1
  \right),
\end{align}
and the
Maharanobis distance, 
\begin{align}
  D_{\rm M}^2
  = 
  \sum_{\ell}^{\ell_{\rm rec,max}}
  \frac{(C_{\ell}^{\rm rec}-C_{\ell}^{\rm org})^2}%
  {\sigma_{\ell}^2},
\end{align}
where $C_{\ell}^{\rm rec}$ is estimated from either a 
pseudo-fluctuation or the IHE reconstructed fluctuations.
Fig. 5 shows the averaged power spectrum of the fiducial simulations
normalized by input spectrum.
We obtained
$\delta C=-0.29$ and
$0.022$ for the pseudo- and IHE reconstructions
and found that for
$\ell\le 5$,
$D^2_{\rm M} = 0.88$ and $0.48$, respectively.
Therefore,
the IHE method provides a less biased estimate
of
the power
spectrum $C_{\ell}$.
On large scales, the suppression of the power due to the masking is mitigated
by the IHE reconstruction, while on smaller scales,
$\ell \simeq \ell_{\rm rec,max}$
and the reconstruction slightly overestimates the power spectrum.

\begin{figure}
  \includegraphics[width=\linewidth]{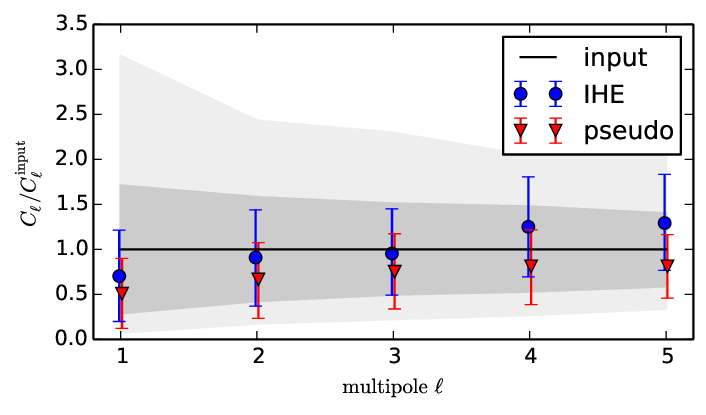}
  \caption{
    Power spectrum reconstruction accuracy for fiducial
    simulation sets. Shown are reconstructed power
    spectra $C_\ell$ normalized by spectrum from original map. 
    Error bars are 1$\sigma$ regions computed from 1000 realizations,
    and shaded regions are 1 and 2 $\sigma$ regions for original 
    spectrum.
    \label{fig:cl}}
\end{figure}

\subsection{Dependence on underlying power spectrum}
\label{ssec:dependence}
The reconstruction accuracy depends on the underlying power spectrum
$C_\ell\propto \ell^{n_s}$ of the fluctuation because the harmonic
modes are not independent in the masked incomplete sky even if the
underlying fluctuation is Gaussian.  We consider three power law
indices, $n_s=0, -2$ and $-4$, which correspond to the following three
cases.  The projected two-dimensional galaxy or dark matter
distribution is approximated as $C_{\ell}^{\rm g} \propto \ell^0$
\cite[e.g.][]{Frithetal:05} on large scales, and the ordinary
Sachs--Wolfe spectrum is approximated as
$C_{\ell}^{\rm SW}\propto \ell^{-2}$ \citep{SachsWolfe:67}.  On very
large--angular scales, the integrated Sachs--Wolfe effect, which gives
$C_{\ell}^{\rm ISW}\propto \ell^{-4}$ \cite[e.g.][]{Cooray:02}
dominates the CMB in the standard $\Lambda$CDM scenario.

Given the typical scale of the mask $\theta_{\rm M}$, it is impossible
to reconstruct a fluctuation whose scale is smaller than
$\ell_{\rm max} \simeq \ell_{\rm M} \ge 180/\theta_{\rm M}$.  Due to
the mode coupling, fluctuations with angular scales corresponding to
$\ell_{\rm M}$ are strongly affected by fluctuations with smaller
angular sizes.  If the spectral index is negative, the amplitude of a
smaller scale fluctuation is weak and does not strongly disturb the
large-scale fluctuations.  Therefore, the deconvolution mapping $F$ is
less affected by singularities.  However, if the spectrum is flat or
has a positive slope, the large-scale modes are highly contaminated by
the small-scale fluctuations.

In \Fref{fig:svd}, the top and bottom panels show the $D^2$ accuracies
for $n_s=0$ and $n_s=-4$. For the $n_s=0$ case, the SVD has a minimum
at $\lambda_k\sim 0.5$ that is larger than when $n_s=-2$.  Thus,
$F_{\rm SVD}(n_s=0)$ is more affected by singularities than
$F_{\rm SVD}(n_s=-2)$.  The IHE results in the right panels shows that
$N_{\rm I}=1$ gives the highest accuracy and that the reconstructed
accuracies gradually degrade to the value given by the brute--force
method.  Conversely, $F_{\rm SVD}(n_s=-4)$ is not affected by
singularities because it always shows
$\partial D^2 / \partial \lambda_k < 0$ for the SVD method.

\subsection{Reconstructed accuracies for each $\ell$ and $m$ mode}
\label{ssec:lmmode}
\begin{figure*}
  \includegraphics[width=\linewidth]{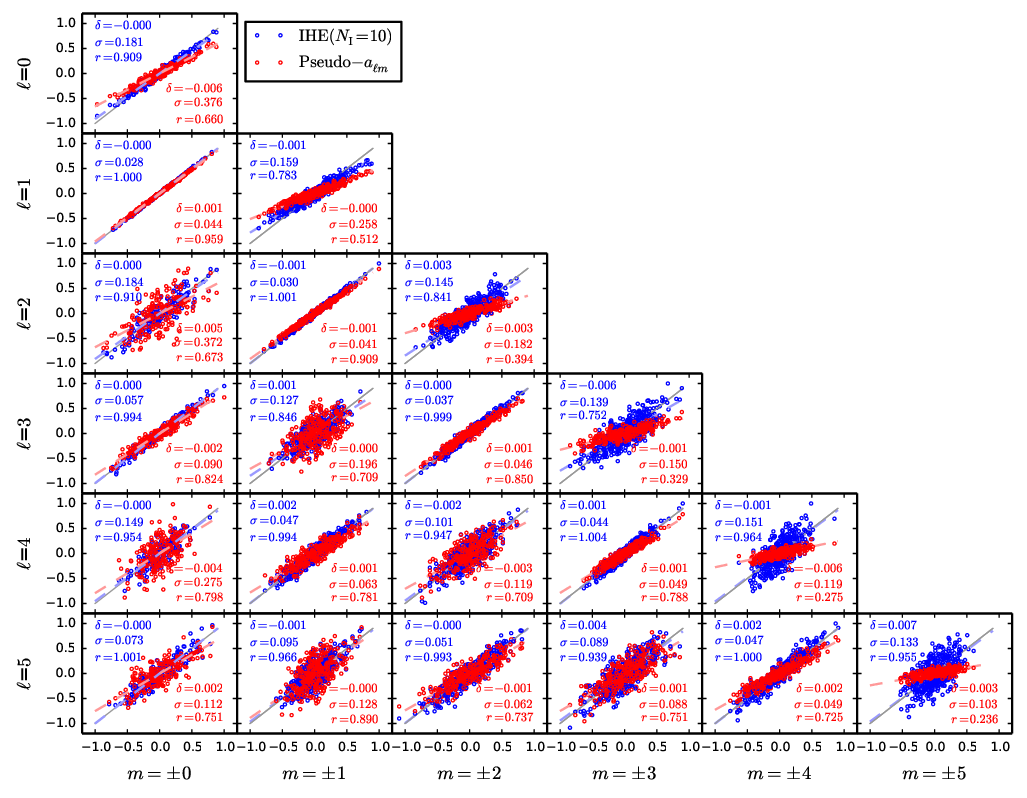}
  \caption{
    Reconstruction accuracy for individual $\ell$ (top to bottom) and $m$
    modes (left to right). We assume that $|b|<20^{\circ}$ region
    is masked and that the underlying density fluctuation obeys
    isotropic Gaussian statistics with the Zel'dovich spectrum
    ($n_s=-2$). In each panel, horizontal axis shows input $a_{\lm}$s
    and vertical axis represents reconstructed $a_{\lm}$s. Each point
    describes one realization. Red and blue
    points are the pseudo--$a_{\lm}$s and $a_{\lm}$s reconstructed
    with $N_{\rm I}=10$ using the IHE method, respectively. For
    illustrative purposes, we show only 100 samples which were
    randomly picked from 1000 realizations, and whose $a_{\lm}$
    magnitudes range from $-1$ to $1$.
    \label{fig:alms}
  }
\end{figure*}
It is important to pay attention to the dependence of the
reconstruction accuracy on the multipoles $\ell$ and $m$.  If there
are particular modes that do not suffer from the masking effect, we
can use them to perform robust cosmological analysis.

\Fref{fig:alms} shows a scatter plot of the original and reconstructed
$a_{\lm}$s for an azimuthally symmetric $\pm 20^{\circ}$ mask. The
solid line represents a relation in which the reconstructed $a_{\lm}$
is identical to the original one.  The dashed red and blue lines in
each panel are the best linear fits for the pseudo--$a_{\lm}$s and the
IHE method, respectively.  The statistical accuracy is also shown in
each panel: $\delta$, $\sigma$ and $r$ are the average differences
between the original and reconstructed $a_{\lm}$s, the standard
deviation, and the best fitted slope of the linear fit, respectively.
We note the following three characteristics:
\begin{itemize}
\item{} For the odd modes ($\ell+m=2n+1$, where $n$ is an integer), 
the masking effect is 
sufficiently small to enable accurate map reconstruction.
In that case, the
  pseudo--$a_{\lm}$'s are already accurate and the IHE method slightly
  improves the accuracy.
\item{} For the even modes ($\ell+m=2n$, where $n$ is an integer), the masking systematically
  suppresses the amplitude of fluctuations.  For a given $\ell$ mode,
  the suppression is more significant for larger $m$ modes.
\item{} For a given $\ell$ mode, the masking effect is the most significant
  for the $\ell=m$ mode and the effect is more significant for higher
  $\ell$ modes.
\end{itemize}

\begin{figure}
  \begin{center}
    \includegraphics[width=\linewidth]{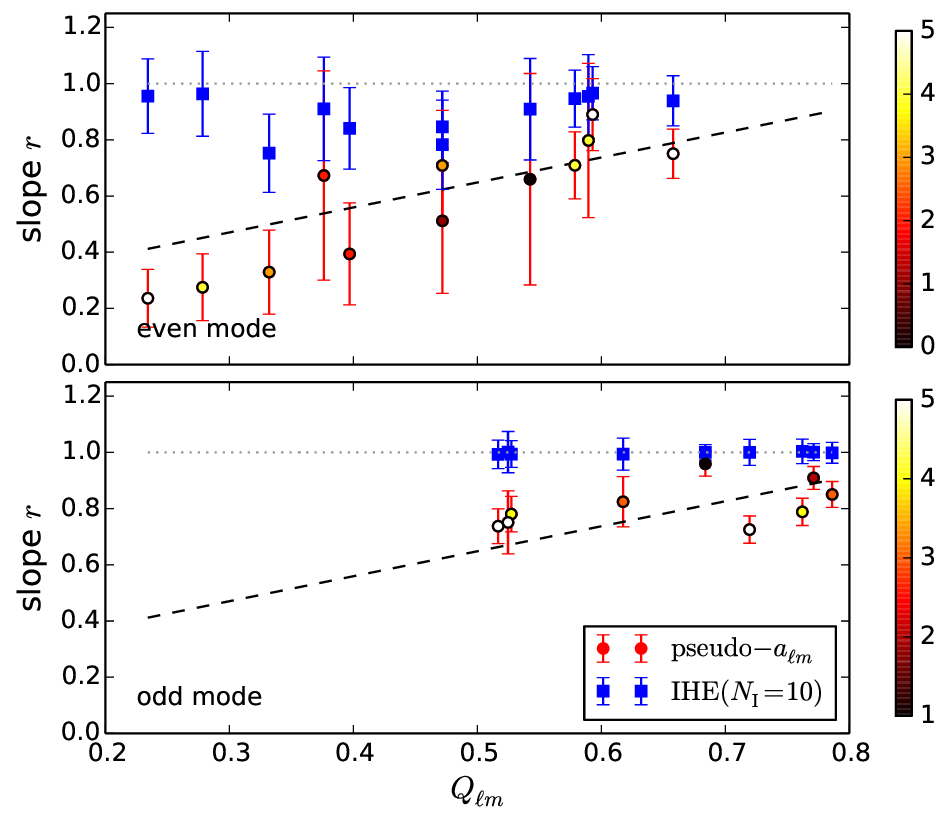}
    \caption{
      Correlation between $Q_{\lm}$ defined in
        \rref{eq:integrated_ylm} and the masking effect probed by the
        fitted slope $r$ for pseudo--$a_{\lm}$s (circle) and IHE
        reconstructed $a_{\lm}$s (square). Colors of circles represent
        corresponding $\ell$ modes shown in color bars at right.
        Error bars are $1\sigma$ regions derived from 1000 random
        simulations. Pseudo--$a_{\lm}$s are linearly correlated with 
        $Q_{\lm}$'s with some errors.
        Top and bottom panels are for even $(\ell+m=2n)$
        and odd $(\ell+m=2n+1)$ modes, respectively.
      \label{fig:Ylm_integral}
    }
  \end{center}
\end{figure}

These dependencies are closely related to the value of the diagonal
part of the mask matrix, i.e.
  \begin{equation}
    \label{eq:integrated_ylm}
    Q_{\lm} \equiv 
    W_{\ell m\ell m} 
    = \int \! \dd\cos\theta \dd\phi 
    \left|
      Y_{\lm}(\theta, \phi) 
    \right|^2
    W(\theta, b),
  \end{equation}
  which quantifies how important the fluctuation outside the masked
  region is and takes values between 0 and 1.  In the limit where the
  {area of the} masked region approaches zero, the mask matrix becomes
  an identity matrix; therefore, $Q_{\ell m} \rightarrow 1$, which
  means that $100\%$ of the fluctuation distribution is outside the
  mask.  For the odd modes, because $Y_{\ell m}$ is small near the
  equator, i.e. within the masked region, the fluctuation in the
  masked region is not significantly different from zero.  In this
  case, $Q_{\ell m}$ is close to unity.  Conversely, for the even
  modes, $Y_{\ell m}$ takes relatively larger values near the equator;
  therefore, the fluctuation inside the mask becomes important
  compared to that outside the mask. In this case, $Q_{\ell m}$ is
  close to zero.  Therefore, the above dependency of the
  reconstruction accuracy is simply correlated with the choice of the
  basis function relative to the mask geometry.  The circle symbols in
  \Fref{fig:Ylm_integral} show the best fit of the slope of the
  pseudo--$a_{\lm}$s obtained in \Fref{fig:alms} as a function of
  $Q_{\lm}$ for the even (top--panel) and odd (bottom--panel) modes,
  respectively.  We can see a clear correlation between the slopes and
  $Q_{\lm}$s.  For the odd modes, the $Q_{\ell m}$s are larger, while
  they are smaller for the even modes.  In each panel, the different
  $\ell$ modes are distinguished by the color levels.  It is evident
  that the lower $\ell$ modes tend to have a larger values of
  $Q_{\lm}$ and, therefore, larger slopes $r$.  The blue squares in
  \Fref{fig:Ylm_integral} are the same as before but for the
  $a_{\ell m}$ reconstructed using the IHE method.  Note that the IHE
  reconstruction works pretty well for the odd modes, as the slope
  $r \sim 1$ implies.  Conversely, the reconstruction for the even
  modes is less accurate than that for the odd modes.

\begin{figure}
  \begin{center}
    \includegraphics[width=\linewidth]{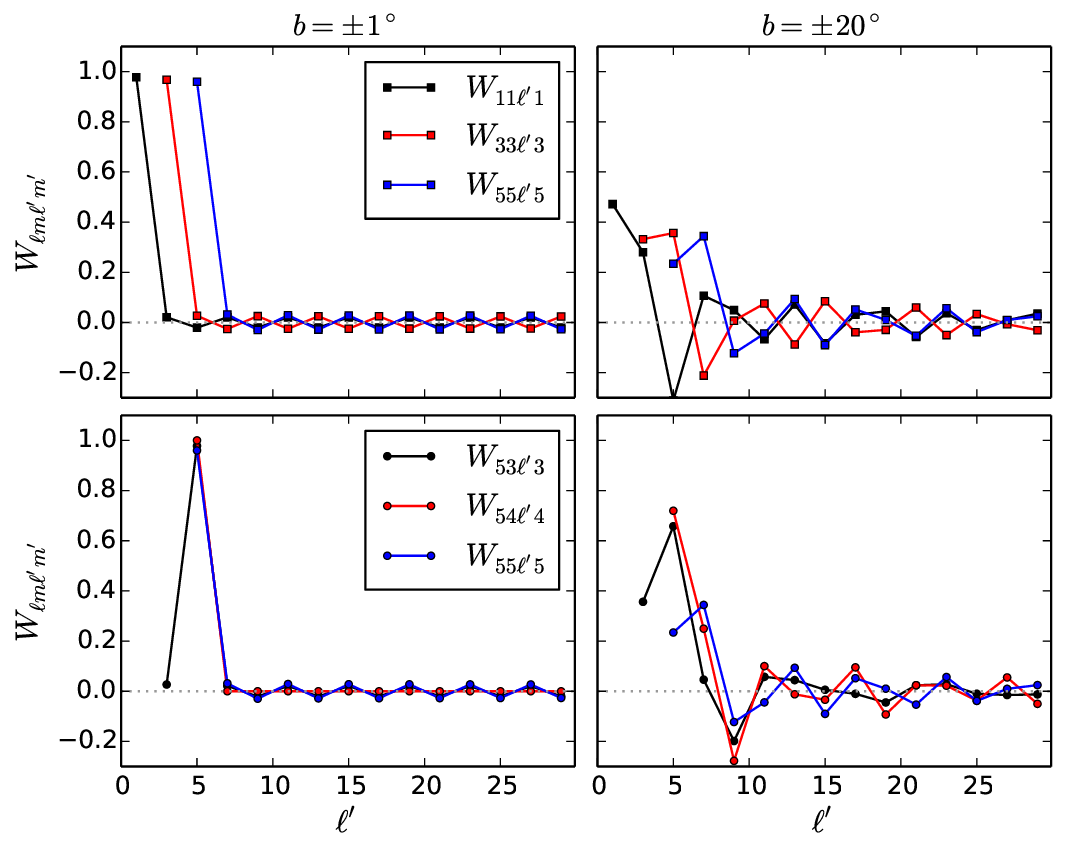}
    \caption{
      Mask matrix, $\bvec{W}$ for (left) $b=\pm 1^{\circ}$
      and (right) $b=\pm 20^{\circ}$. Top and bottom panels correspond
      to $m=m'=\ell$ modes and $m=m', \ell=5$ modes, respectively. 
      Here, we show only non--zero components. For $b=\pm 1^{\circ}$
      case, matrix is close to identity matrix, while for $b=\pm
      20^{\circ}$ case, diagonal parts leak into the off--diagonal
      parts, which induces mode coupling between different $\ell, m$
      modes in $a_{\ell m}$s.
      \label{fig:Wlm}
      }
  \end{center}
\end{figure}

  Even though $Q_{\lm}$ can explain the strength of suppression of the
  $a_{\lm}$s, it does not have a perfect correlation with the slope
  $r$.  Consequently, we need to consider other factors as well.  As
  we mentioned above, $Q_{\ell m}$ represents the amount of
  fluctuations leaking outside the masked region.  In other words,
  $Q_{\ell m}$ represents a fraction of the diagonal components in the
  mode coupling matrix $W_{\ell m \ell' m}$, where
  $\sum_{\ell'} W_{\ell m \ell' m}=1$ for a given $\ell$ and $m$ mode.
  Note that for azimuthally symmetric masks, the
  $W_{\ell m \ell' m'}$s have non--zero values only for the $m=m'$ and
  $\ell' = \ell + 2n$ modes, where $n=0, 1, 2,\cdots$.  \Fref{fig:Wlm}
  shows the non--zero components of the mask matrix,
  $W_{\ell m\ell' m}$.  The top and bottom panels show the $\ell=m$
  modes for different $\ell$ and the different $m$ modes for $\ell=5$,
  respectively. In the top panel, it is evident that $\bvec{W}$ at
  $\ell'=m$, which is $Q_{\ell m}$, decreases monotonically with
  $\ell'$.  It is also apparent that for the $b=\pm 1^{\circ}$ case,
  the matrix is almost diagonal, while for the $b=\pm 20^{\circ}$
  case, there is a long tail towards higher $\ell$ modes. This tail
  can induce mode coupling between different $\ell$ modes, and the
  strength of the coupling depends on each realisation of the map.
  The scattering of points centered at the dashed lines in
  \Fref{fig:Ylm_integral} may be due to this effect.
%

\section{Application to CMB Sky}
\label{sec:realistic}

\subsection{Planck Galactic mask and CMB power spectrum}
\label{ssec:real}
We again consider an isotropic Gaussian prior; however, now, we
discuss the application of the IHE method to more realistic
cases. First, we describe a reconstruction of the CMB map, which is
masked by the Galactic plane.  In the concordant model, on
superhorizon scales, the spectral index of the CMB power spectrum is
approximately given by $n_s=-4$ because the ISW effect is dominant. On
horizon scales, the index increases to $n_s=-2$, which corresponds to
the Harrison-Zel'dovich spectrum because the ordinary Sachs-Wolfe
effect is dominant. On subhorizon scales, the index increases to
$n_s \simeq -1$ if the scale is smaller than the sound horizon at the
last scattering time.

\begin{figure}
  \includegraphics[width=\linewidth]{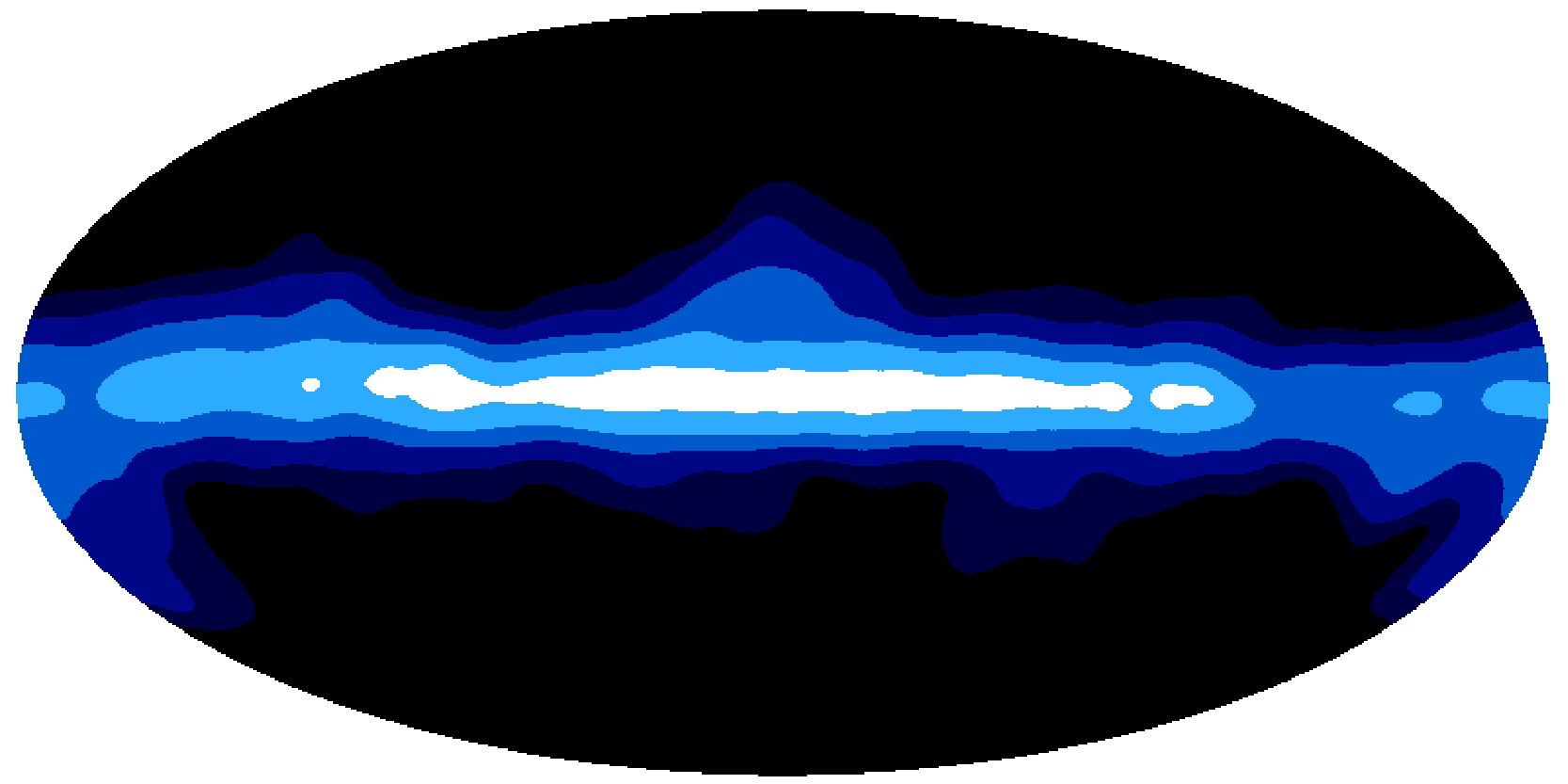}
  \caption{Planck Galactic plane masks without
    apodization: from black to white, GAL60, 70, 80, 90 and 97
    respectively. 
    \label{fig:mask}}
\end{figure}

In this study, we used the Galactic masks provided by the Planck DR2
\footnote{http://irsa.ipac.caltech.edu/data/Planck/release\_2/ancillary-data/}
\citep{PlanckXI:2015},
and neglected the point source masks because the masks for each point
source were too small to affect the reconstruction of the large mode
fluctuations, $\ell<10$.  However, note that the fluctuations inside
the point source mask could be well reconstructed by limiting the
reconstruction area in the vicinity of the point source mask region
instead of the entire sky. In that case, we could redefine a local
orthogonal system, i.e. the harmonics in the finite two-dimensional
flat space. However, such analysis was beyond the scope of this study
and will be investigated in the future.  Figure \ref{fig:mask}
illustrates the Galactic masks we used. The colors (from black to
white) show different masking schemes in which the areas of the
unmasked regions are $60\%, 70\%, 80\%, 90\%$ and $97$ \% of the
entire sky.  We generated 1000 random Gaussian simulations with an
input spectrum from the latest Planck $\Lambda$CDM cosmological model
\cite{Planck2015:cosmology}.

\begin{figure*}
  \begin{tabular}{ccc}
    \includegraphics[width=0.33\linewidth]{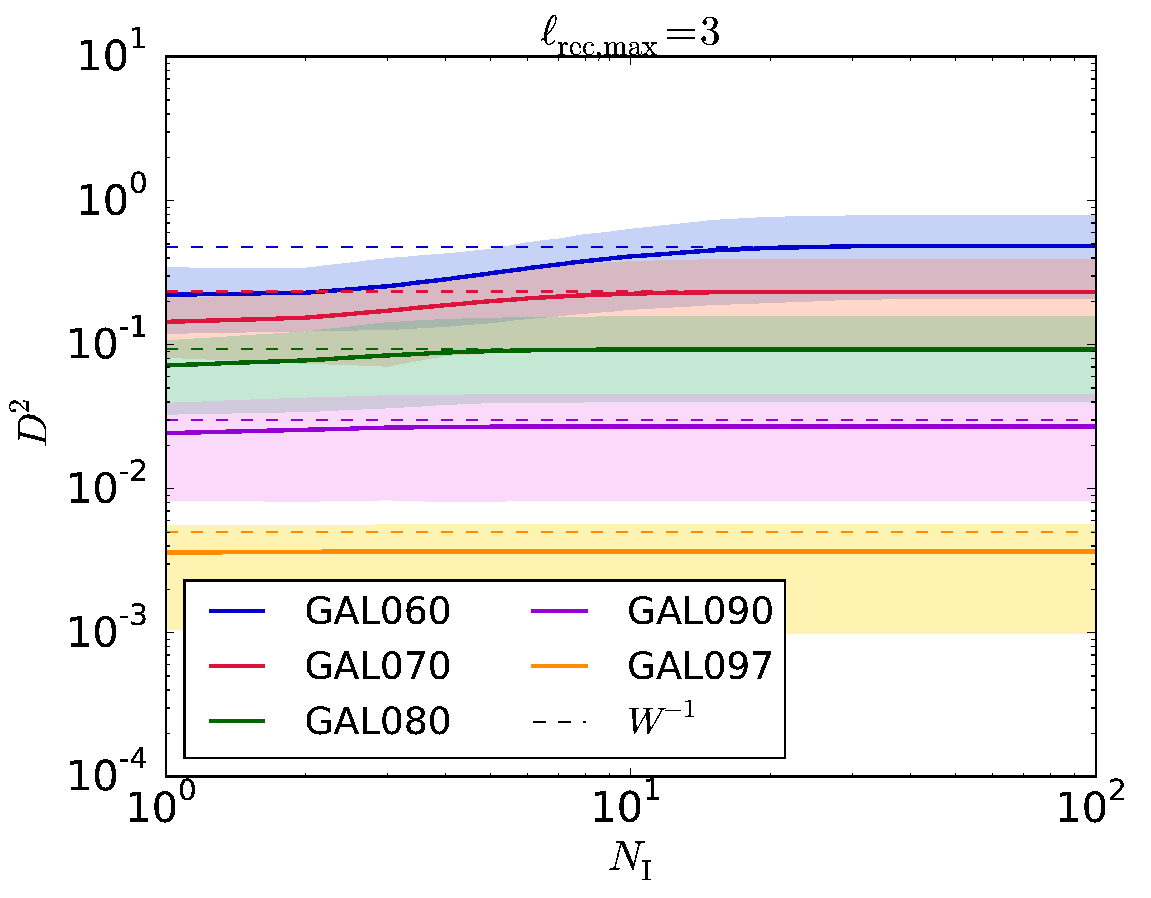} &
    \includegraphics[width=0.33\linewidth]{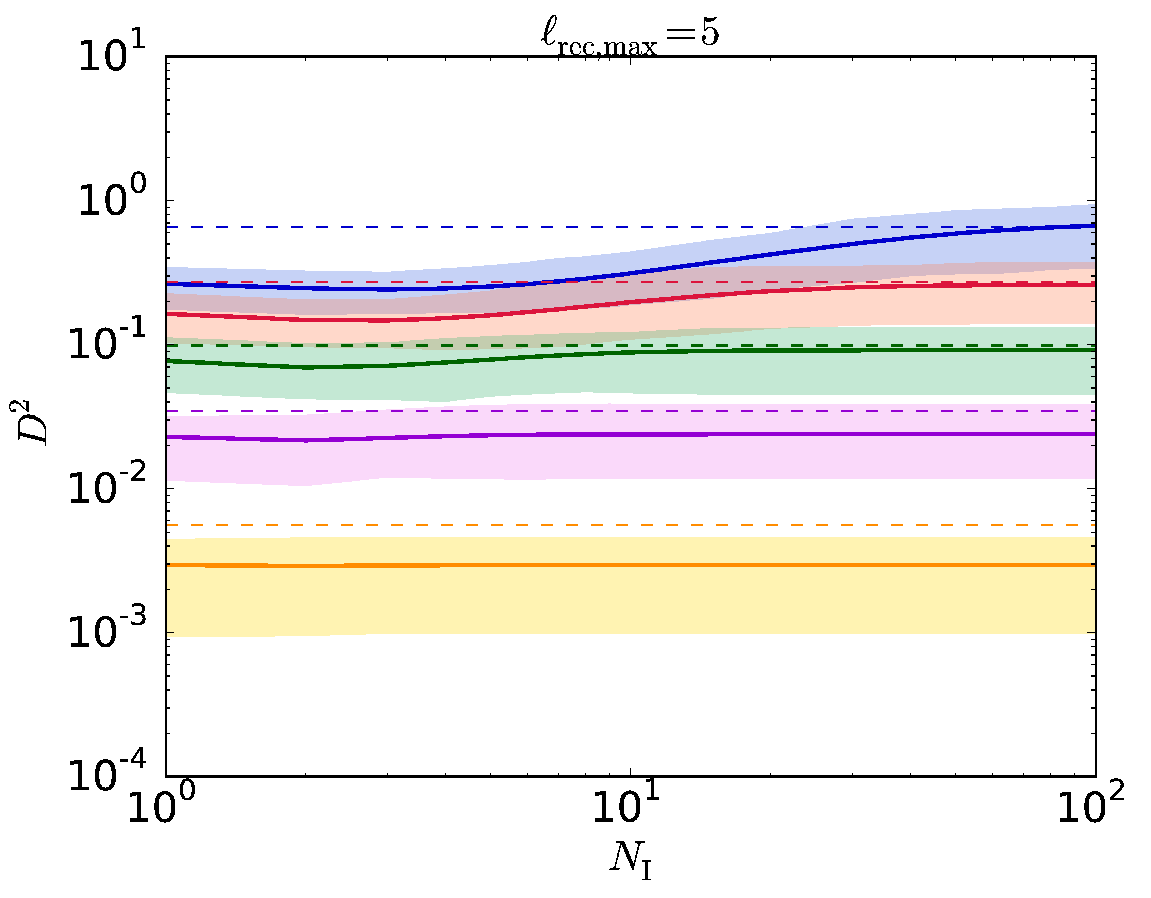} &
    \includegraphics[width=0.33\linewidth]{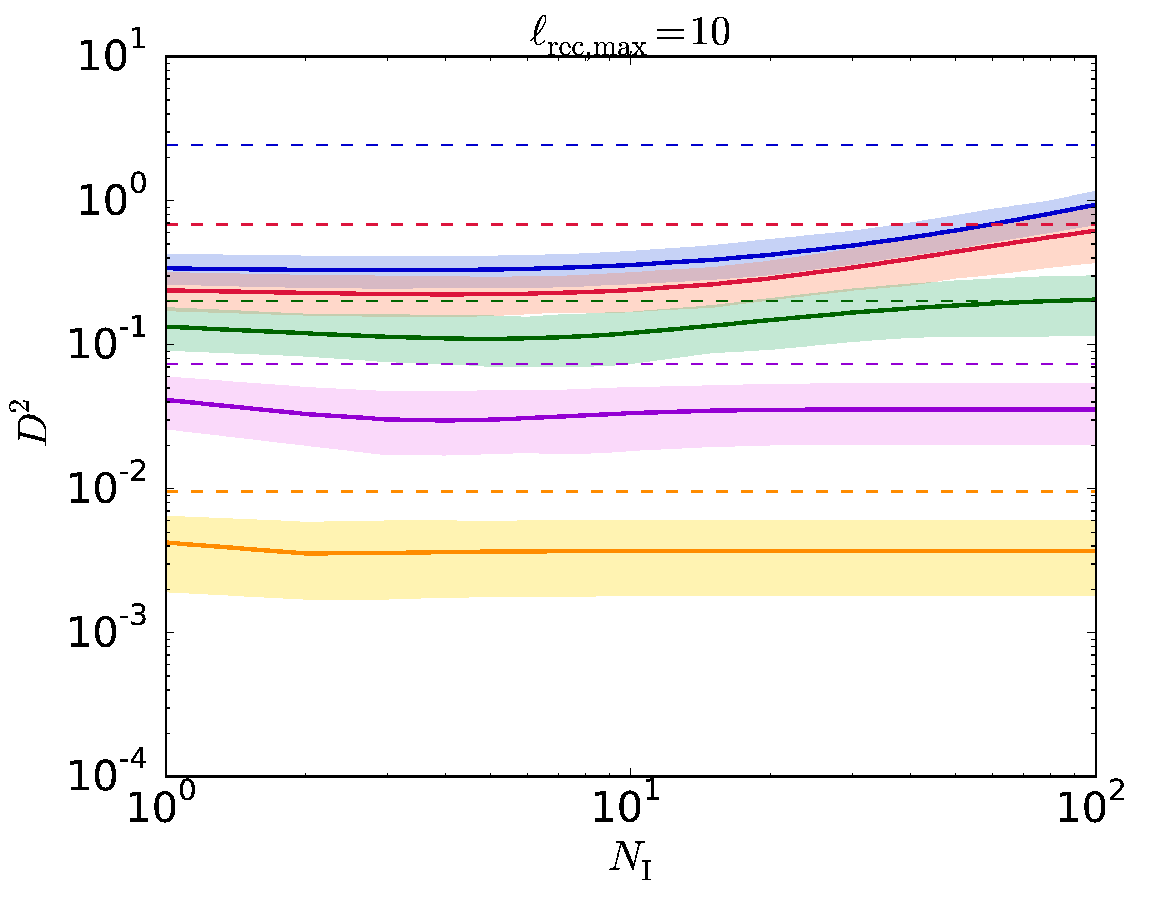} \\
  \end{tabular}
  \caption{
    Reconstruction accuracy for simulated CMB maps masked by Galactic
    plane. Colors from top to bottom indicate different masks shown in
    inset. For each color, solid line marks central value and  shaded
    region enclosed by thin lines indicates 1 $\sigma$ region for the
    1000 random realizations.
    \label{fig:cmb}}
\end{figure*}

Figure \ref{fig:cmb} 
presents
the reconstruction accuracy for the
simulated CMB maps masked by the Galactic plane. 
For all the $\ell_{\rm rec,max}$, the IHE reconstruction with a finite
number of iterations gives better reconstructions than the direct inversion.
By comparing
the results with those shown in Figure \ref{fig:svd}, 
remembering
that
$65$\% 
of the sky of
a $b=20$ mask is unmasked,
it can be seen that
the $n_s=-2$ 
results are
consistent with the case in which the Galactic
plane is masked.
As 
mentioned, the detailed shape of the
mask does not significantly affect the reconstruction 
using the IHE method if the
area of the masked sky is similar.
The optimal number of iterations depends on the size of
the mask and the
reconstruction scale $\ell_{\rm rec,max}$; therefore,
the number of iterations should be determined
before reconstructing the CMB map.

\subsection{Non-Gaussian Prior}
\label{ssec:nG}
In the 
past sections,
we 
demonstrated
that the IHE reconstruction method
could
be applied to 
Gaussian isotropic fluctuations. In practice, the map may contain
non-Gaussian features. In this section, we consider two types of
non-Gaussian priors. First, we consider a type of isotropic
non-Gaussian
fluctuation
induced by
primordial non-Gaussianity in the density perturbation. 
The local type non-Gaussianity on large-scale 
CMB fluctuations
can be written as,
\begin{equation}
  \frac{\Delta T}{T}
  = 
  \left( \frac{\Delta T}{T} \right)_{\rm  G} 
  -3 f_{\rm NL}
  \left\{
  \left( \frac{\Delta T}{T} \right)_{\rm  G}^2 - 
  \left \langle \left( \frac{\Delta T}{T} \right)_{\rm  G}^2\right\rangle  \right\},
\end{equation}
where $\langle \Delta T/T \rangle_{\rm G}$ is a Gaussian fluctuation
of the CMB temperature in the sky.  The map is scaled to its root mean
square to be $\langle \Delta T/T \rangle_{\rm G}=10^{-4}$ so that
readers can compare the value of the $f_{\rm NL}$ to the one
introduced in the CMB analysis. We show the reconstruction accuracies
for $f_{\rm NL}=0$ and $5000$ for comparison.  Given that the recent
CMB observation by Planck suggest that the value of $f_{\rm NL}$ is
consistent with zero \citep{Planck2015NG}, the value of
$f_{\rm NL}=5000$ assumed here might be too large.  However, even in
such an extreme case of non-Gaussianity, we found that the IHE
reconstruction accuracy was not affected much.  Here, we fixed the
azimuthal mask size to $b=20^\circ$ as before.

\begin{figure}
  \includegraphics[width=\linewidth]{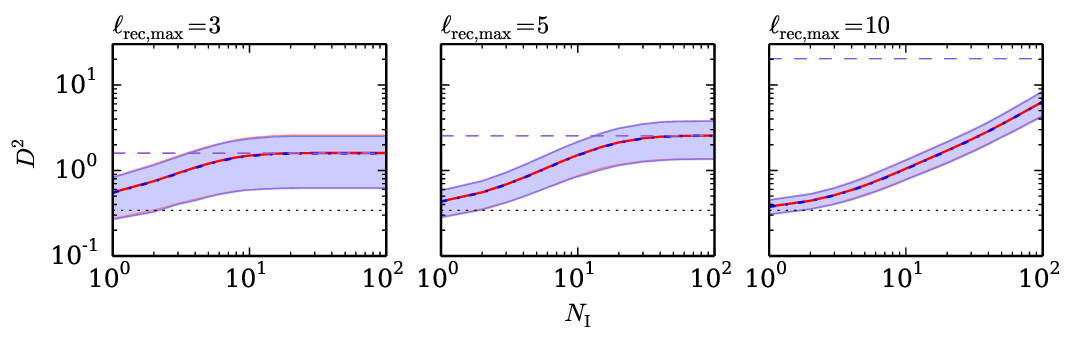}
  \includegraphics[width=\linewidth]{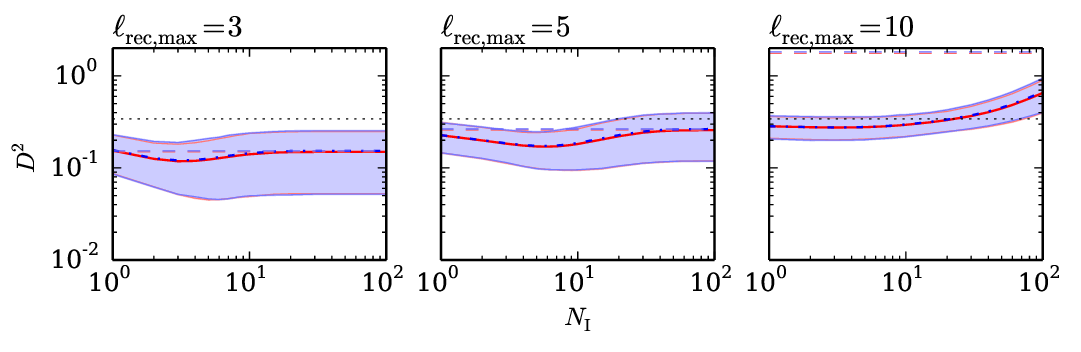}
  \includegraphics[width=\linewidth]{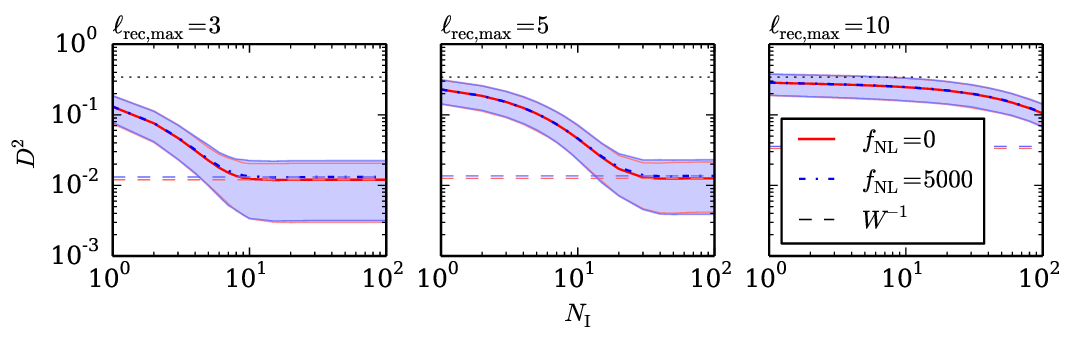}
  \caption{
    Comparison of the map reconstruction accuracy between Gaussian
    (red solid) and non-Gaussian (blue dashed) fluctuations. 
    From top to bottom, the spectra of the Gaussian fluctuations are
    $n_s=0, -2$ and $-4$ respectively. Even
    for large non-Gaussianity, $f_{\rm NL}=5000$, IHE recovers
    same reconstruction accuracies for all $\ell_{\rm
      rec,max}$ and $n_s$.
    \label{fig:ngfnl}}
\end{figure}

Figure \ref{fig:ngfnl} shows the map reconstruction accuracies as a
function of the number of iterations for IHE. The results are compared
with the isotropic Gaussian prior case.  It is evident that even if
the non-Gaussianity is quite large, such as $f_{\rm NL}=5000$, the
reconstruction accuracy does not significantly change compared to the
isotropic Gaussian prior case. This result implies that the IHE method
is robust against the probability distribution of the underlying
fluctuation as long as the statistical isotropy holds.

Second, we consider an anisotropic non-Gaussian prior. For instance,
if there is a large non-Gaussian structure in the universe, it can
affect the CMB sky via the ISW effect. To simulate this situation, we
first generated a Gaussian isotropic map and then added a circular
structure that had a Gaussian radial profile:
\begin{equation}
  \frac{\Delta T}{T}(\uv{n})
  =
  \left( \frac{\Delta T}{T} \right)_{\rm  G} (\uv{n})
  +
  A'_\sigma \exp\left[ \frac{(\uv{n}-\uv{n}_0)^2}{2 r_\sigma^2} \right],
\end{equation}
where $A'_\sigma, r_\sigma$ and $\uv{n}_0=(\theta_0, \phi_0)$ are free
parameters that specify the amplitude, size and the center position of
a non-Gaussian structure.  $A'_\sigma$ is the amplitude of a circular
symmetric non-Gaussian structure with a Gaussian profile with a radius
$r_\sigma$.

It is more convenient to rewrite the parameter as
$A'_\sigma = A_\sigma \ell(\ell+1)C_\ell/2\pi$, where
$\ell = \pi/r_\sigma$.  We chose to use $A_\sigma = 1$ and $10$ and
$r_\sigma=10^{\circ}$ and $40^{\circ}$ in this study.  Note that the
azimuthal position of the structure $\phi_0$ does not affect the
results because the background Gaussian field is statistically
isotropic and the mask we considered is independent of the azimuthal
position.  Therefore, we can always set $\phi_0=0$ without losing
generality.  $\theta_0$ was chosen such that the center of structure
would correspond to the center of the mask
($\theta_0=90 [\textrm{deg}]$), the edge of the mask
($\theta_0=70 [\textrm{deg}]$), and completely outside the mask
($\theta_0=0 $).

\begin{figure}
  \includegraphics[width=\linewidth]{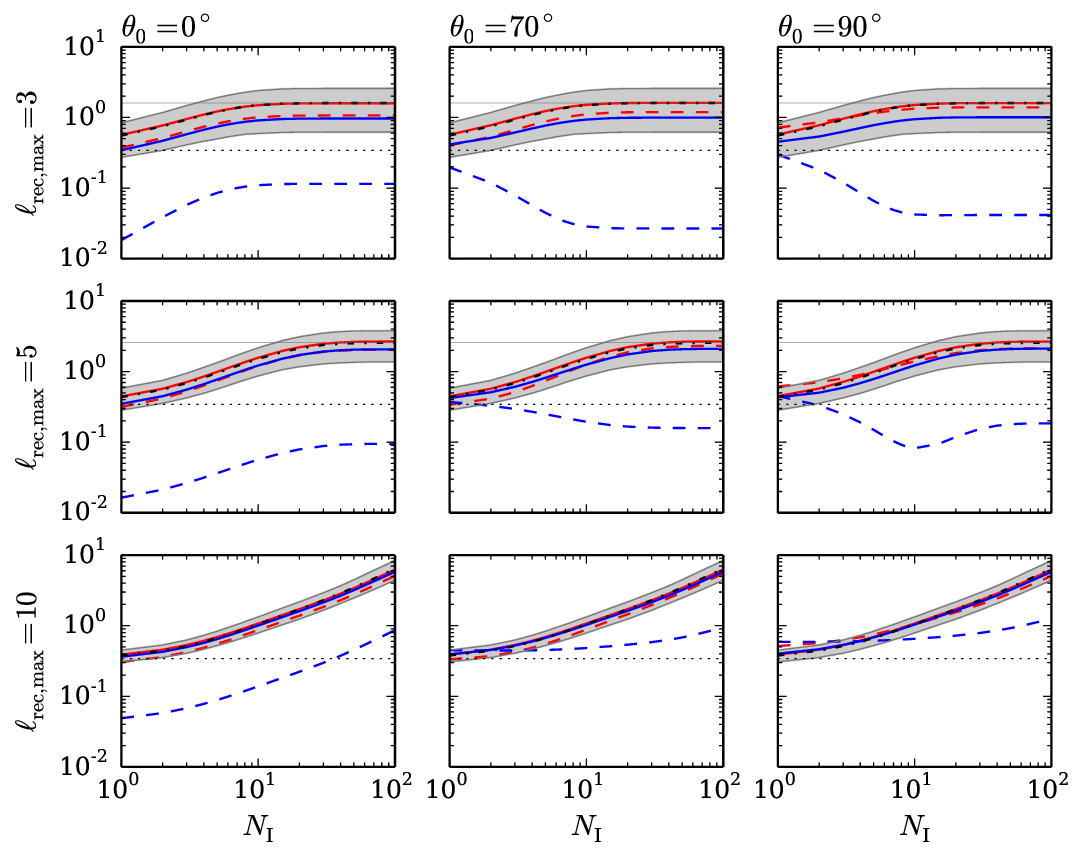}
  \caption{Reconstruction accuracy of anisotropic non-Gaussian maps
    for $n_s=0$. From left to right, center of non-Gaussian structure
    is located at $\theta_0=0^{\circ}, 70^{\circ}$ and
    $90^{\circ}$. From top to bottom, reconstruction multipoles are
    $\ell_{\rm rec,max}=3, 5$ and $10$, respectively. Different lines
    indicate different sizes and amplitudes structure shown in inset
    of \rref{fig:ning_blue}.  Gaussian cases are denoted with black
    dot-dashed lines together with 1 $\sigma$ error regions from 1000
    random simulations.
    \label{fig:ning_red}}
\end{figure}

\begin{figure}
  \includegraphics[width=\linewidth]{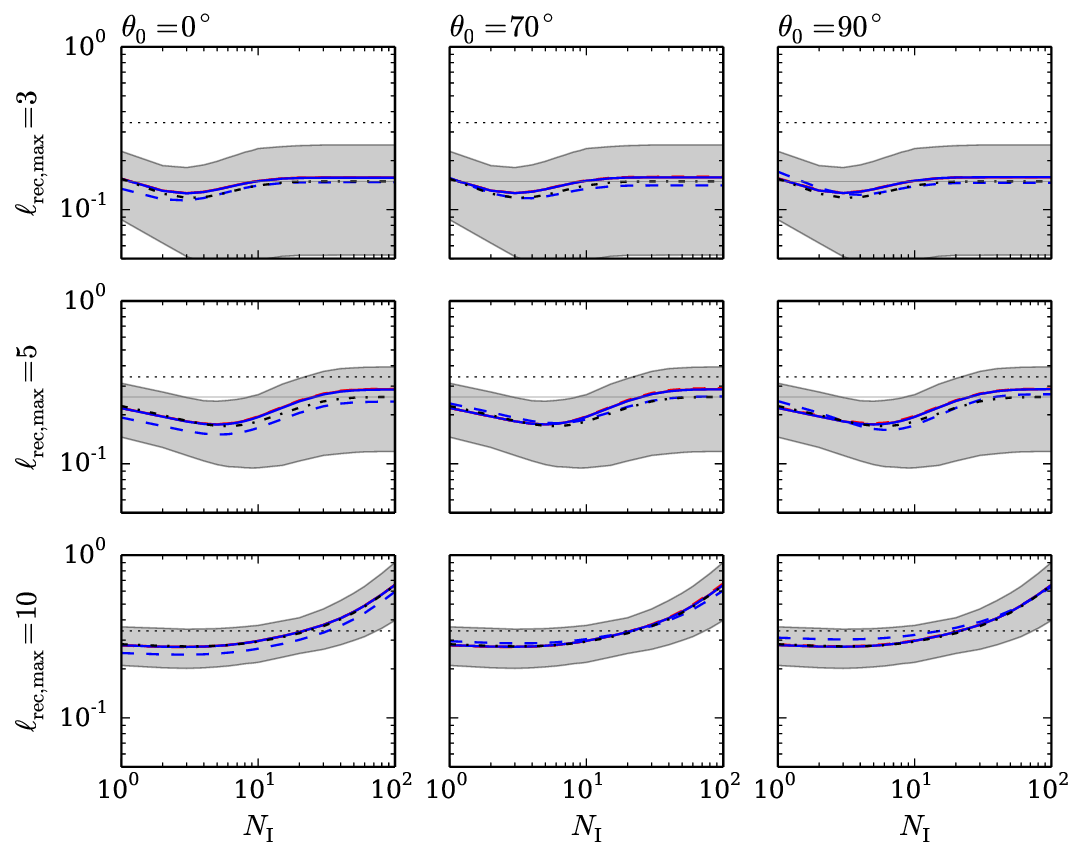}
  \caption{Same as Fig. \ref{fig:ning_red} but for $n_s=-2$
    background spectrum.\label{fig:ning_white}}
\end{figure}

\begin{figure}
  \includegraphics[width=\linewidth]{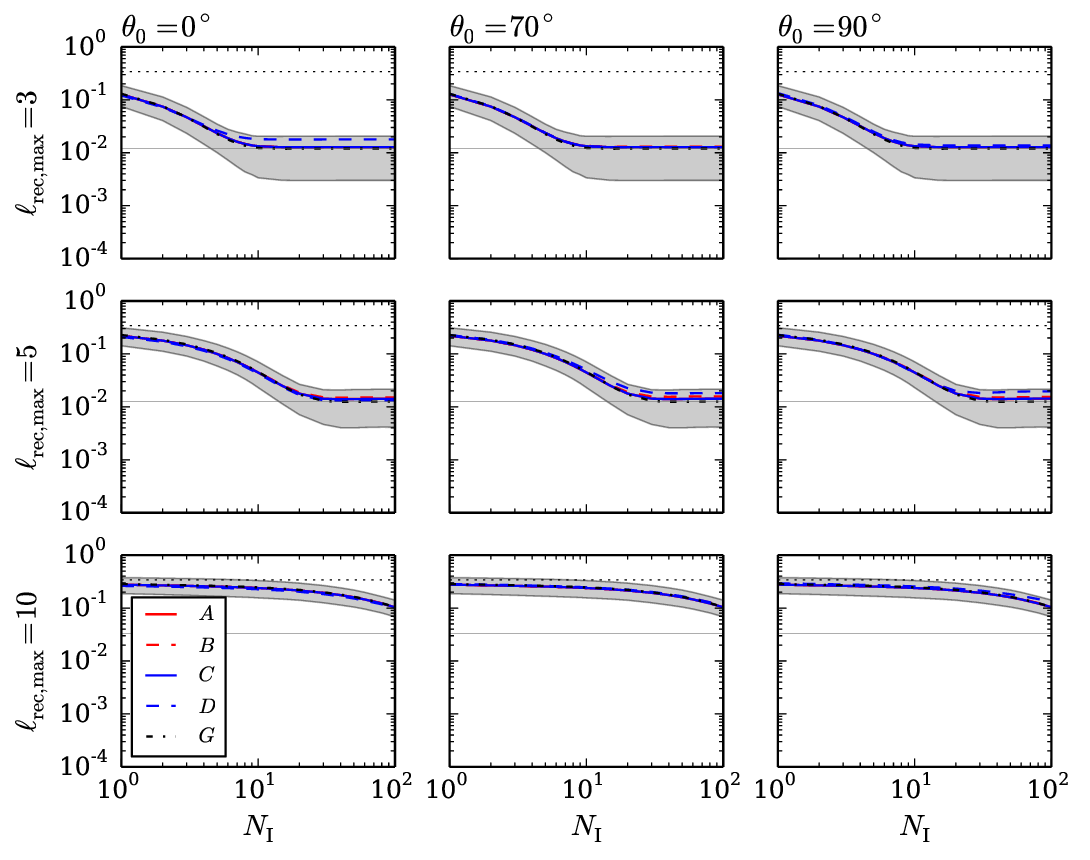}
  \caption{Same as Fig. \ref{fig:ning_red} but for $n_s=-4$
    background spectrum. 
    Red solid and dashed lines (denoted as $A$ and $B$ in the
      inset) show cases
      where $r_{\sigma}=10$ and $A_{\sigma}=1$
      and $10$, respectively. Blue solid and dashed lines (denoted as 
      $C$ and $D$) show cases where
      $r_{\sigma}=40$ and $A_{\sigma}=1$ and $10$, respectively.
      Black dashed line is case with only Gaussian fluctuation.
    \label{fig:ning_blue}}
\end{figure}

Figures \ref{fig:ning_red}-\ref{fig:ning_blue} show the map
reconstruction accuracies as a function of the number of iterations
for background fluctuations with power-law indices $n_s=0, -2$ and
$-4$.  For the $n_s=-2$ and $-4$ backgrounds, the effect of the
non-Gaussian structure is very small, while for $n_s=0$ background,
the effect is prominent.  We obtained higher reconstruction accuracies
for larger structures.  Therefore, adding a structure with $r_\sigma$
increases the amplitude of the power spectrum at
$\ell \simeq \pi/r_\sigma$, e.g. $\ell \sim 3$ for
$r_\sigma = 40^{\circ}$.  For the $n_s=0$ spectrum, adding power to
large scales may greatly change the effective slope of the spectrum so
that the spectrum becomes redder, which acts to mitigate the
mode--mode coupling. However, further studies are necessary to examine
the reason for this phenomenon.


\section{summary}
\label{sec:summary}
We investigated the map reconstruction accuracy with the IHE for
isotropic Gaussian fluctuations and isotropic and anisotropic
non-Gaussian fluctuations as well as the realistic CMB fluctuation
when the sky was masked near the Galactic plane.  Reconstructing the
missing data in a masked region is known as an inverse problem.  We
found that the IHE method is equivalent to brute--force inversion in
the limit that the number of iterations approaches infinity.  However,
in particular cases, finite truncation of the iterations results in a
better estimate of the underlying fluctuation.  The reconstruction
accuracy depends on the size of the mask $b$, the maximum multipole
mode to be reconstructed $i_{\rm rec,max}$ and the spectral index of
the underlying fluctuation $n_s$ The IHE method is equivalent to the
asymptotic expansion in terms of the mask size $b$ and it converges to
the correct values in the limit of $b\rightarrow 0$.

As an example, we applied the IHE method to reconstruct the data
obscured by azimuthally symmetric masks. We considered three types of
Gaussian fluctuations with power--law indices of $n_s=0, -2$ and $-4$,
which correspond to the matter or galaxy power spectrum, the ordinary
Sachs--Wolfe spectrum and the integrated Sachs--Wolfe spectrum,
respectively, in the context of cosmological analyses.  For the
$n_s=-2$ case, we found that there exists an optimal finite number of
iterations that makes the reconstruction more accurate than the SVD
method or the brute--force matrix inversion method.  For the $n_s=0$
case, the pseudo--$a_{\lm}$ is the best estimator of the projected
density fluctuations.  For the $n_s=-4$ case, the brute--force
inversion method yields the highest accuracy.  In that case, the IHE
method can help reduce the computation time for inversion.

We also found that for azimuthally symmetric masks, the amplitudes of
the reconstructed fluctuations for the even ($\ell+m=2n$) modes are
significantly suppressed in comparison to the odd modes
($\ell+m=2n+1$).  For a fixed $\ell$ mode, the $m=\ell$ mode is more
affected by the masking than by other $m \ne \ell$ modes, and the
suppression is more prominent for higher $\ell$ modes.  Therefore, the
IHE method reproduces odd modes more accurately.  The suppression due
to masking can be explained by the deviation of $Q_{\lm}$ from unity;
however, the strength of the mode coupling that changes at each
realisation may also affect the suppression in a complex manner.

We demonstrated that the IHE method can be applied to reconstruct
realistic CMB observations.  For large-scale modes, $\ell<10$, the IHE
method provides more accurate reconstructed maps than direct inversion
does, and the optimal number of iterations should be determined before
reconstructing the CMB.  For some special cases, we investigated the
IHE reconstruction accuracy for both isotropic and anisotropic
non-Gaussian fluctuations.  For isotropic non-Gaussian fluctuations,
which are characterized by $f_{\rm NL},$ the reconstruction is not
substantially affected by non-Gaussianity, which only changes the
amplitude of the power spectrum but does not affect its tilt.  As an
example of anisotropic non-Gaussianity, we added a single structure
with a Gaussian radial profile onto an isotropic background Gaussian
fluctuation.  For the $n_s=0$ spectrum, adding such a non-Gaussian
structure dramatically improves the reconstruction accuracy compared
to the isotropic Gaussian case, while for the $n_s=-2$ and $-4$
spectra, the effect of non-Gaussianity is negligible.

It would be interesting to investigate how the significance of the
large--angle CMB anomaly changes when we use different methods of map
reconstruction.  We will explore this problem in our future work.

\section*{Acknowledgments}
We thank Masahiro Takada, Eiichiro Komatsu, Issha Kayo and Takahiro 
Nishimichi for the useful discussions. 
AN was supported in part by the FIRST program ``Subaru Measurements of Images and Redshifts
(SuMIRe)'', CSTP, Japan. 
This work was also supported in part by
MEXT KAKENHI Grant Number 16H01096.

\bibliographystyle{mn2e}
\bibliography{bibdata}

\end{document}